\documentclass[a4paper,fleqn]{cas-dc}
\usepackage[numbers]{natbib}
\usepackage[utf8]{inputenc}
\usepackage{amsmath}
\usepackage{soul}
\usepackage{cuted}

\date{\today}

\usepackage{graphicx}
\usepackage{color}
\usepackage{amssymb}

\begin{document}
\let\WriteBookmarks\relax
\def\floatpagepagefraction{1}
\def\textpagefraction{.001}

\shorttitle{Viscous dissipation in smooth sliding motion}
\shortauthors{SV Sukhomlinov et~al.}

\title[mode = title]{On the viscous dissipation caused by randomly rough indenters in smooth sliding motion}

\author[1,2]{Sergey Sukhomlinov}[orcid=0000-0001-8759-9367]

\author[1,2]{Martin H. Müser}[orcid=0000-0003-0919-0843]\ead{martin.mueser@mx.uni-saarland.de}
\cormark[1]

\address[1]{Dept. of Materials Science and Eng., Saarland University, 66123 Saarbrücken, Germany}
\address[2]{INM -- Leibniz Institute for New Materials, Campus D2 2, 66123 Saarbrücken, Germany}

\cortext[cor1]{Corresponding author}

\begin{abstract}
    The viscous dissipation between rigid, randomly rough indenters and linearly elastic counter bodies sliding past them is investigated using Green's function molecular dynamics.
    The study encompasses a variety of models differing in the height spectra properties of the rigid indenter, in the viscoelasticity of the elastomer, and in their interaction.
    All systems reveal the expected damping linear in sliding velocity $v$ at small $v$ and a pronounced maximum at intermediate $v$.
    Persson's theory of rubber friction, which is adopted to the studied model systems, reflects all observed trends.
    However, close quantitative agreement is only found up to intermediate sliding velocities.  
    Relative errors in the friction force become significant once the contact area is substantially reduced by sliding. 
\end{abstract}


\maketitle

\section{Introduction}
In 2001, Bo Persson, who is honored in this special issue of 
Applied Surface Science Advances, published an article with the title \textit{Theory of rubber friction and contact mechanics}~\cite{Persson2001JCP}.
The importance of this paper to tribology in general and to contact mechanics in particular can barely be overestimated.
It is the first theoretical approach to the description of contacts between nominally flat surfaces making reliable predictions on many interfacial properties possible, at least in the important limiting case of linearly elastic bodies.
The aspect of this seminal work on static contact mechanics has been scrutinized with many rigorous large-scale simulations.
Agreement is found to be generally excellent, in particular for the dependence of mean separation on normal pressure~\cite{Persson2007PRL,Yang2008JPCM,Almqvist2011JMPS,Campana2011JPCM,Pastewka2013PRE,Dapp2014JPCM,Muser2017TL} and the gap distribution function as well as the leakage rate that follows from it~\cite{Dapp2012PRL}, including the leakage rate close to the percolation threshold~\cite{Dapp2016SR} and for anisotropic surfaces~\cite{Persson2020EPJE,Persson2021TL,Wang2021TL}. 

One of the appeals of Persson's theory is the ease with which it can be applied and extended to interfaces other than the default system consisting of a semi-infinite, elastic body in repulsive contact with an isotropic, randomly rough counterface. 
Often, it is sufficient to identify the correct expression for how the elastic energy depends on the wave vector of a sinusoidal surface undulation in full contact to address an entirely new contact problem. 
Comparison between Persson's theory and accurate numerical approaches include the analysis of adhesion with half spaces~\cite{Carbone2009EPJE-adh,Persson2014JCP,Joe2017TI,Wang2017TL} and thin elastic plates~\cite{Carbone2004PRB}, anisotropic roughness~\cite{Carbone2009EPJE-ani,Wang2021TL}, and generalized (graded) elastic manifolds~\cite{Muser2021TL}, to name a few.  

In contrast to the many, just-mentioned tests on the validity of Persson's theory regarding static contact mechanics, to which Persson's own successful contribution to the contact mechanics challenge~\cite{Muser2017TL} can be added, the capability of his theory to describe viscous dissipation induced by the sliding motion of randomly rough indenters past elastomers has been scrutinized surprisingly little with stringent numerical methods.
In fact, despite significant progress in numerical boundary-value simulations of visco-elastic solids
~\cite{Pastewka2012PRB,Carbone2013JMPS,Carbone2014PRE,Putignano2015IJSS,Scaraggi2015JPCM,Kajita2016PRE,Bugnicourt2017TI,Bugnicourt2018TL,Menga2018TI,Putignano2019JMPS}
over the last decade, only few studies
~\cite{Scaraggi2015JPCM,Bugnicourt2017TI,Menga2018TI,Afferrante2019EPJE}
presented a direct
comparison of numerically rigorous simulations to Persson's theory.
In our perceiption, those latter works address predominantly a narrow
parameter range with a focus on large relative contact areas,
where good agreement with theory is found~\cite{Scaraggi2015JPCM,Bugnicourt2017TI,Menga2018TI,Afferrante2019EPJE}, but
contain
inconclusive results on the friction at velocities large enough to
substantially reduce the contact area.
For example, only one~\cite{Afferrante2019EPJE} of the studies that we are aware of tested Persson's theory by comparing its predictions to numerically accurate reference data~\cite{Putignano2019JMPS} in a way that we find most meaningful, that is,
by studying how the friction (coefficient) depends on velocity $v$ at a constant normal pressure,
for which the static ($v = 0$) relative contact area is clearly less than
one half.
While Afferrante \textit{et al.}~\cite{Afferrante2019EPJE} find semi-quantitative agreement between theory and simulations, even at large velocities when contact areas are small, others~\cite{Scaraggi2015JPCM,Bugnicourt2017TI,Menga2018TI} find agreement only at relative contact areas $a_\textrm{r} \gtrsim 0.5$.
However, when the sliding velocity is large and $a_\textrm{r} \lesssim 0.2$,
deviations between theory and simulations appear to be large, as can be seen,
for example,
in Fig.~{2} of Ref.~\cite{Scaraggi2015JPCM},
in Figs.~{8-11} of Ref.~\cite{Bugnicourt2017TI},
and
in Fig.~8 of Ref.~\cite{Menga2018TI}.
Moreover, none of the works incorporate inertial effects, which, in
principle, are easily encoded into Persson's theory, as we demonstrate in
this study, but generally defy those approaches~\cite{Carbone2013JMPS,Carbone2014PRE,Putignano2015IJSS} assuming the Green's function
to factor into a time-dependent and a spatially dependent function.

To investigate the validity of Persson's rubber friction theory more comprehensively than before, we simulate the sliding motion of randomly rough indenters past elastic counterfaces.
In this endeavor, we replace the commonly made  non-overlap constraint with a repulsive interaction, in which the energy density increases quadratically with the overlap between the surfaces.
This choice, which contains the non-overlap constraint as a limiting case, makes it possible to extend our favourite numerical technique for contact-mechanics simulation, namely, Green's function molecular dynamics (GFMD)~\cite{Campana2006PRB}, from quasi-statics to dynamics.
At the same time, Persson's theory is readily adapted to account for such finite microscopic contact stiffness making it possible to meaningfully compare theory and simulation. 
In addition, we adjust his theory to reflect inertial effects, which are included in some of the employed viscoelastic models. 
We also extended our house-written GFMD code such that the standard-linear-solid model in the Kelvin-Voigt representation, which Persson used in his original work on rubber friction~\cite{Persson2001JCP}, could be simulated. 

The remainder of this article is organized as follows:
In Sect.~\ref{sec:modelANDmetods}, we introduce the used models and methods.
Sect.~\ref{sec:theory} contains a brief summary of Persson's rubber friction theory including our modification to make it account for intertial effects and finite-range repulsion. 
Results are presented in Sect.~\ref{sec:results}, while conclusions are drawn in Sect.~\ref{sec:conclusions}. 

\section{Model and Methods}
\label{sec:modelANDmetods}

In this section, we present our model for the sliding contact of a (hypothetical) elastomer past a randomly rough indenter.
To this end, we separate the model description into three parts, namely, the structural properties of the indenter, the viscoelastic properties of the elastomer, and the interaction between the counterfaces.
Once the model is set up, it is straightforward to implement the features into our house-written GFMD code, which has been described numerous times before~\cite{Campana2006PRB,Prodanov2014TL,Zhou2019PRB}.
Two new, important features were added to the code for this work: 
first, the use of the standard-linear-solid model and second, the way how sliding is imposed and lateral forces are measured. 
These  aspects are discussed in separate subsections.
Finally, we present simulations of the retraction from and the sliding motion past a flat punch in this section to illuminate the dynamical properties of the various elastomers.

\subsection{Randomly rough indenter}
The height spectrum of a randomly rough indenter is generally assumed to cross over smoothly between being constant at small wave numbers $q$ and to a power law dependence in $q$ at large $q$, according to~\cite{Majumdar1990W,Palasantzas1993PRB,Persson2014TL,Jacobs2017STMP}
\begin{equation}
    C(q) = 
    \frac{C_\textrm{r}\,\mathrm{\Theta}(q_{\textrm{s}}-q)}
    {\left(1 + q^2/q_\textrm{r}^2 \right)^{1+H}}
    \label{eq:rollOff}
\end{equation}
Here, $C_\textrm{r}$ is the height spectrum at the roll-off wave number $q_\textrm{r}$, $\Theta(...)$ denotes the Heaviside step function, $H$ is called the Hurst exponent, and $q_\textrm{s}= 2\pi/\lambda_\textrm{s}$ is the wave vector associated with the short wavelength cutoff $\lambda_\textrm{s}$.

In addition to the default, smooth roll-off spectrum
we also consider a hard cut-off spectrum, defined through
\begin{equation}
    C(q) = C_\textrm{r}\, 
    ({q_\textrm{r}}/{q})^{2(1+H)}\,
    \mathrm{\Theta}(q_{\textrm{s}}-q)\,
    \mathrm{\Theta}(q-q_{\textrm{r}}).
\end{equation}
It allows some analytical results to be obtained more easily than for realistic spectra, which is why it is a useful reference from a theoretical perspective.
As a compromise between smooth roll off and hard cut off, we also
use a hard roll off spectrum, in which
$C(q\le q_\textrm{r}) = C_\textrm{r}$, while 
for other $q$, the spectrum is identical to that used for a hard cutoff.

When defining a surface, the Fourier coefficient of the height $h(\mathbf{r})$ is set to $\tilde{h}(\mathbf{q}) = \sqrt{C(q)}\,\exp(2\pi\,i u_\mathbf{q})$, where $u_\mathbf{q}$ is an independent random number that is uniformly distributed on $(0,1)$.
Of course, $\tilde{h}(0)$, which is nothing but the center-of-mass height of the indenter surface, is not assigned a random variable. 
Instead it is chosen such that the highest point of the surface equals zero. 

As default values, we use
$\lambda_\textrm{r} = 0.4~L$,
$\lambda_\textrm{s} = 0.004~L$, and
$H = 0.8$.
The specific value of $C(q_\textrm{r})$ is irrelevant for this study, as we assume linear elasticity and report all results  in reduced units so that no single reported number depends on the specific value of $C(q_\textrm{r})$. 
However, we mention for completeness that heights are always normalized in our code such that the mean-square height gradient equals unity. 

\subsection{Viscoelastic properties of the elastomer}
\label{sec:cGFMD}

The viscoelastic properties of a linearly elastic solid determine its dynamical response to an external time-dependent stress, whereby they define  the equations of motion. 
In a reverse conclusion, it can be argued that the way how the equations of motion are solved define the viscoelastic properties of the \textit{in-silico} solid.
In this sense, the default GFMD dynamics, which are set up as to let the simulation quickly relax to static solutions, scarcely reflect realistic dynamics.
They are nevertheless well defined dynamics, which allow us to ascertain to what extent Persson's theory faithfully reflects inertial effects.
This is why we consider GFMD dynamics in addition to dynamics reflecting standard, linearly viscoelastic solids, which we solve in a similar way as Bugnicourt \textit{et al.}~\cite{Bugnicourt2017TI}, who also used a Fourier based approach to model viscoelastic half spaces.

Finally, the small-slope approximation is assumed. 
This concerns both, the viscoelastic aspects as well as the interfacial interactions. 

\subsubsection{Conventional GFMD dynamics}
In most Fourier-based GFMD simulations conducted so far, the equation of motion of a surface mode $\tilde{u}(\textbf{q})$ satisfies
\begin{equation}
    m_q \,\ddot{\tilde{u}}(\textbf{q},t) + \gamma\, m_q\, \dot{\tilde{u}}(\textbf{q},t) + \frac{qE^*}{2} \,\tilde{u}(\textbf{q},t) = \tilde{\sigma}(\textbf{q},t),
\end{equation}
where $m_q$ is the inertia associated with a given mode,
$\gamma$ is the rate with which velocity is damped, and $\tilde{\sigma}(\mathbf{q},t)$ is the Fourier transform of the stress acting on the solid's surface, which includes interfacial and external stresses.
In this work, the latter is simply the nominal pressure $p_0$ acting on the center-of-mass mode.

Here and in the following, $\sigma$ is meant to refer to compressive stresses.
In our calculations,
the elastomer is squeezed down with the nominal pressure $p_0$ against the substrate.
This way, a positive stress leads to a positive displacement and the mean interfacial stress is equal to $p_0$ in steady-state sliding. 
When presenting results visually, we found it more intuitive to revert that set-up. 

The dynamical properties of the model are defined by the choice of the inertia $m_q$ and the damping $\gamma$, the latter of which could also depend, in principle, on the wave vector.
We define a reference mass $m_\textrm{ref}$ such that if $m_\textrm{ref}$ were assigned to the stiffest mode, the intrinsic frequency of the stiffest mode of the system, $\tilde{u}(q_\textrm{max})$, would be unity in an appropriate unit system, i.e.,
\begin{equation}
    m_\textrm{ref} = \frac{q_\textrm{max}\,E^*}{2}\,
    [t]^2,
\end{equation}
where $[t]$ is the unit of time in our unit system. 
Two different approaches are used in this work:
In the original, or, regular GFMD~\cite{Campana2006PRB}, the inertia of all modes is chosen equal, while in  mass-weighting (MW) GFMD~\cite{Zhou2019PRB}, an attempt is made to collapse the frequencies at small relative contact area, by assigning smaller inertia to longer wavelength undulations. 
While the proportionality in MW-GFMD is usually made a function of the normal pressure $p_0$ with which the elastic solid is squeezed down on the indenter, we decided to use an inertia that does not depend on $p_0$, to ease the interpretation of numerical results.

To summarize, the following two choices were made
\begin{equation}
m_q = m_\textrm{ref} \times 
\begin{cases}
1 & \textrm{regular GFMD} \\
\sqrt{q^2_0+q^2}/q_\textrm{max} & \textrm{MW-GFMD},
\end{cases}
\end{equation}
where $q_0 = 2\pi/L$ is the smallest non-zero wave vector fitting into the periodically repeated, square simulation cell having edges of length $L$. 
Finally, the damping term is chosen as
\begin{equation}
    \gamma = [t]^{-1}
    \begin{cases}
    q_0/q_\textrm{max} & \textrm{regular GFMD} \\
    1 & \textrm{MW-GFMD}
    \end{cases}
\end{equation}
such that the slowest internal modes are close to being critically damped in regular GFMD, i.e., modes with wave number $q_0$.
In MW-GFMD, all modes are reasonably close to critical damping~\cite{Zhou2019PRB}.

\subsubsection{Standard-linear-solid dynamics}
\label{sec:sls}

\begin{figure}[hbtp]
    \centering 
    \includegraphics[width=0.5\textwidth]{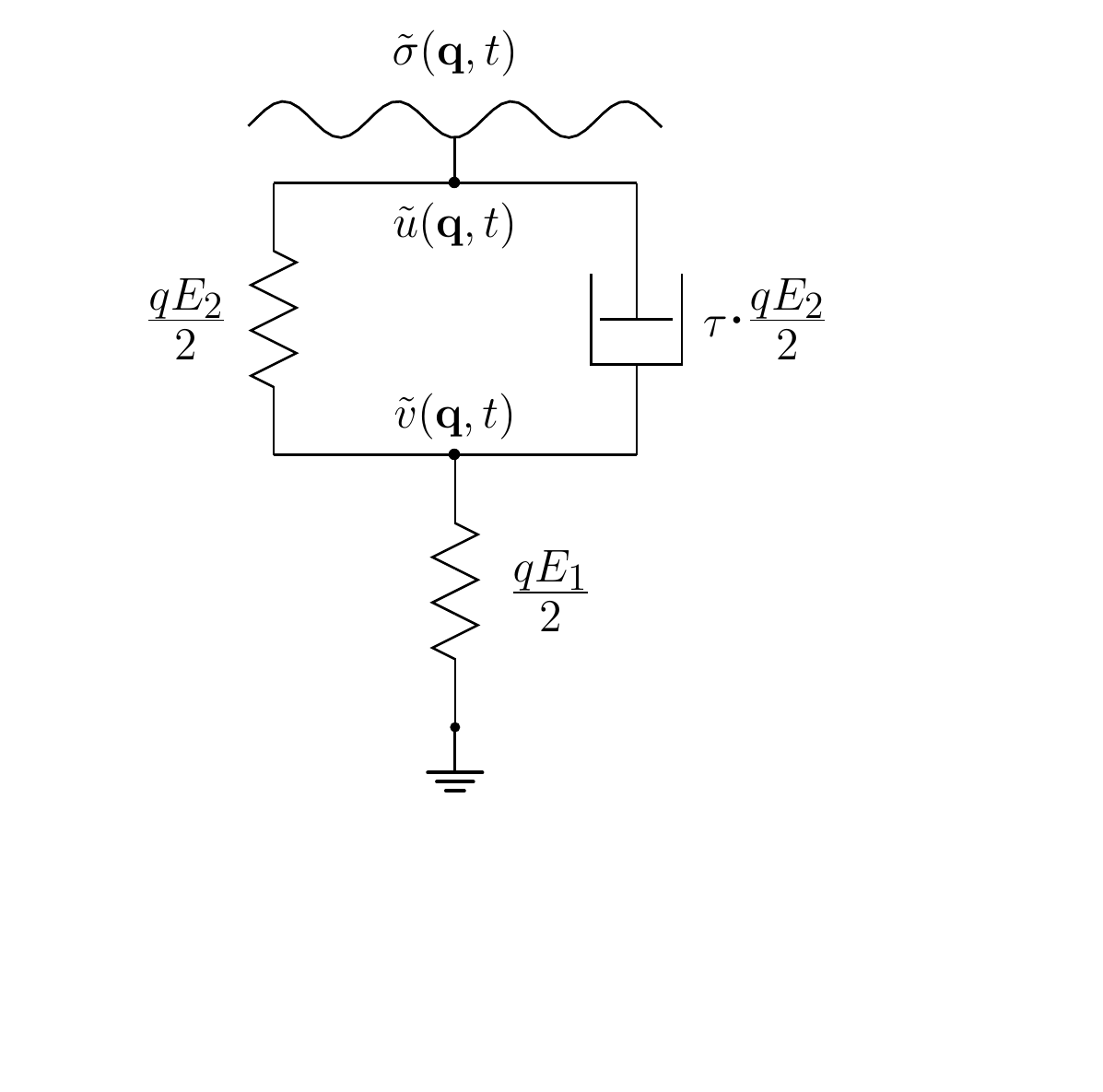}\vspace*{-1.5cm}
    \caption{Standard linear solid model in the Kelvin-Voigt representation with the parameters used in the GFMD simulations.}
    \label{fig:SLSmodel}
\end{figure}

To obtain more realistic dynamics than those produced by conventional GFMD,  we also consider the standard model for linearly elastic solids in the Kelvin-Voigt representation.
It had also been used in Persson's pioneering paper on rubber friction~\cite{Persson2001JCP}.
The model contains two degrees of freedom and is depicted in Fig.~\ref{fig:SLSmodel}.

In the notation of Sect.~\ref{sec:cGFMD}, the equations of motion read: 
\begin{strip}
\begin{subequations}
  \begin{eqnarray}
\gamma\, m_q\, \dot{\tilde{u}}(\textbf{q},t) + \frac{qE_2}{2} \,\tilde{u}(\textbf{q},t)  & = & 
\gamma\, m_q\, \dot{\tilde{v}}(\textbf{q},t) + \frac{qE_2}{2} \,\tilde{v}(\textbf{q},t) + 
\tilde{\sigma}(\textbf{q},t) \;\;\;\;\;\;\;\\
\gamma\, m_q\, \dot{\tilde{v}}(\textbf{q},t) + \frac{q(E_1+E_2)}{2} \,\tilde{v}(\textbf{q},t) & = & \gamma\, m_q\, \dot{\tilde{u}}(\textbf{q},t) + \frac{qE_2}{2} \,\tilde{u}(\textbf{q},t).
 \end{eqnarray}
 \end{subequations}
\end{strip}
Substituting $\gamma\,m_q$ with $\tau\, q\,E_2/2$ and realizing that
$\tilde{v}(\mathbf{q},t) = 2 \, \tilde{\sigma}(\textbf{q},t) / (q\,E_1)$, the equation of motion for $\tilde{u}(\mathbf{q},t)$ simplifies to~\cite{Bugnicourt2017TI}
\begin{eqnarray}
\frac{q}{2} \left\{
\tau\, \dot{\tilde{u}}(\mathbf{q},t) + \tilde{u}(\mathbf{q},t)\right\} =
\frac{E_1+E_2}{E_1\,E_2}\, \tilde{\sigma}(\textbf{q},t)
+ \frac{\tau \,\dot{\tilde{\sigma}}(\mathbf{q},t)}{E_1}  ,
\label{eq:SLSfinalEOM}
\end{eqnarray}
from where it can be easily deduced that the static contact modulus, $E^* = E_1\,E_2/(E_1+E_2)$, results from a series coupling of two static compliances.

Obviously, Eq.~(\ref{eq:SLSfinalEOM}) cannot be used for the centre-of-mass mode, $u_0(t) \equiv \tilde{u}(0,t)$, because the prefactor to $\dot{u}_0(t)$ disappears.
We therefore replaced the prefactor to the $\dot{u}_0(t)$ term with $q_\textrm{CM}^\textrm{eff}E_2\tau/2$ and treated $q_\textrm{CM}^\textrm{eff}$ as a free parameter, whose value only affects the stability of the integration scheme as well as how quickly $u_0(t)$ approaches its steady-state value.
If we had simulated the relative sliding motion of two rough bodies, its proper choice would necessitate knowledge of body heights, $h$.
To lowest order, it should be possible to set $q_\textrm{CM}^\textrm{eff} = q_0\,L/h$, at least as long as $h$ is not (distinctly) less than $L$.
If this were not the case, the prefactors to the elastic restoring forces have need to be changed anyway. 

A final extended note on our numerical solution of Eq.~(\ref{eq:SLSfinalEOM}) is in place.
We estimate $\dot{\tilde{\sigma}}(\textbf{q},t)$ with a numerical first-order finite difference of the current and the previous value of ${\tilde{\sigma}}(\textbf{q},t)$.
This induces a systematic lag of half a time step in $\dot{\tilde{\sigma}}(\mathbf{q},t)$, which could certainly be alleviated through a predictor method. 
However, a predictor increases the demands on memory and computing time.
More importantly, it deteriorates the numerical stability when sudden events occur or when the substrate potential is very stiff.
This would become particularly relevant if short-range adhesion were included.
In those cases, the displacement field jumps between two or even several branches from one time step to the next. 
Reducing the time step by as much as a factor of 100 did not generally alleviate the situation, even when jumps occurred only between two branches. 
To stabilize the integration scheme in such situations, we applied a low-pass filter to the (original) r.h.s. of Eq.~(\ref{eq:SLSfinalEOM}) of the type $f_\textrm{lpf}(t) = (1/\tau)\,\int_0^t\!\mathrm{d}t'\,f(t')$, where $f(t)$ is the original r.h.s., while $f_\textrm{lpf}(t)$ is the function obtained after low-pass filtering, which is then used as new r.h.s. of Eq.~(\ref{eq:SLSfinalEOM}).
Of course, it needs to be insured that $\tau$ is small compared to any characteristic time scale in the system. e.g., small compared to the discretization length over sliding velocity. 
An alternative solution to the stability issue could be the use of an implicit integration method, which, however, would be quite demanding on memory, computing time, and coding time, while not necessarily promising success.  
In contrast, realizing a low-pass filter necessitates only one additional, large array to be allocated and a few minutes of coding.

\subsection{Interfacial interaction}
When elastomer and indenter have a positive gap, $g(\textbf{r})$, their interaction energy is set to zero. 
Once they start overlapping, i.e., once $g(\textbf{r})<0$, their energy density increases quadratically with the overlap.
Thus, the total interfacial energy in our model reads
\begin{equation}
    V_\textrm{int} = \frac{\kappa_\textrm{i}}{2} 
    \int d^2 r\, g^2(\textbf{r})\,  \Theta(-g(\textbf{r})),
\end{equation}
where $\Theta(...)$ is the Heaviside step function. 

The default choice of the stiffness parameter is
\begin{equation}
\kappa_\textrm{i}^\textrm{D} = 0.2 q_\textrm{max}E^*,
\end{equation}
where $q_\textrm{max}$ is the maximum wave number contained in the simulation cell, i.e., $q_\textrm{max} = \sqrt{8}\pi\,n_x/L$, where $n_x$ is the number of grid points into which the elastomer is discretized parallel to one spatial direction. 
This way, the interfacial potential is close to being as stiff as the stiffest mode in the system, i.e., $\kappa_\textrm{i}$ is roughly as large as possible without creating the need of having to (substantially) reduce the time step compared to a simulation of a free but initially strained surface. 
To test the relevance of this choice for final results and to demonstrate that Persson's theory was generalized correctly for finite interfacial stiffnesses, $\kappa_\textrm{i}$ was also set to values different than its default. 
To approach the non-overlap constraint, we also consider a local interfacial contact stiffness of $\kappa_\textrm{i}=4\,\kappa_\textrm{i}^\textrm{D}$.
This latter choice led to dynamical instabilities, which we believe were predominantly caused from the $\dot{\tilde{\sigma}}(\mathbf{q},t)$ term on the rhs of Eq.~\eqref{eq:SLSfinalEOM}.
To eliminate this unintended behavior, we implemented the low-pass filter for the elastic stresses mentioned at the end of Sec.~\ref{sec:sls}, which stabilized all modes.

\subsection{Imposing sliding and measuring lateral forces}
To impose sliding without (significant) discretization artifacts, we store the initial Fourier coefficients of the height profiles and propagate them in time according to
\begin{equation}
    \tilde{h}(\textbf{q},t) = 
    \tilde{h}(\textbf{q},0) \, e^{i\, \mathbf{q} \cdot \mathbf{v}\,t}.
\end{equation}
After each time step, the inverse Fourier transform is taken so that the precise height of the sliding indenter is known at the grid points representing the elastomer.

The instantaneous lateral force in the in-plane-direction can be given as 
\begin{equation}
    \mathbf{f} = - \nabla_{\Delta \mathbf{r}} V_\textrm{tot}(\Delta \mathbf{r}),
\end{equation}
where $\Delta \mathbf{r}$ is a virtual, rigid, relative, displacement of the two counter bodies by the in-plane vector $\Delta \mathbf{r}$.  
In the Derjaguin approximation (i.e., assuming interfacial interactions to depend only on the local height difference but neither on height gradients nor on deformation gradients nor on related terms), it can be evaluated to be
\begin{equation}
    \mathbf{f} = -\int_A \!d^2r \, \sigma(\mathbf{r}) \, \nabla h(\textbf{r}) ,
\end{equation}
if the compliant top solid is displaced by $\Delta \mathbf{r}$ relative to the stiff bottom indenter having the height profile $h(\mathbf{r})$.

The magnitude of the instantaneous dissipated power can be obtained by multiplying the force from the damping element times the velocity.
Thus, for conventional GFMD dynamics, the dissipated power is given by
\begin{subequations}\label{eq:disspatedPowerTimeDomainA}
  \begin{equation}
    P_\textrm{d}(t) = A\,\gamma  \sum_\mathbf{q}  m_q \left\vert \dot{\tilde{u}}(\mathbf{q},t) \right\vert^2 
\end{equation}
with $\dot{\tilde{v}}(\mathbf{q},t) = 2 \, \dot{\tilde{\sigma}}(\textbf{q},t) / (qE_1)$ and $A = L^2$ the nominal contact area,  while for SLS dynamics
\begin{equation} \label{eq:disspatedPowerTimeDomainB}
    P_\textrm{d}(t) = 
    \frac{A\,\tau\, E_2}{2} 
    \sum_\mathbf{q}\, q 
\left\vert 
\dot{\tilde{u}}(\mathbf{q},t) - \dot{\tilde{v}}(\mathbf{q},t)
\right\vert^2
\end{equation}
must be used.
\end{subequations}

In steady-state sliding, the mean lateral force corresponds to the friction force whenever there are no external constraints beyond constant sliding velocity, so that the friction force at constant velocity $\mathbf{v}$ is given by:
\begin{equation}
    \mathbf{F}(t) = -\,\frac{P_\textrm{d}(t)}{v}\, \frac{\mathbf{v}}{v}.
    \label{eq:forceFromDissipation}
\end{equation}
Thus, estimates obtained through the direct computation of the lateral force (method~1) and those obtained by the measurement of the dissipated power (method~2) approach each other  during ``running in'', which happens quite rapidly in the considered set-up of a rough indenter sliding past an originally flat but compliant counterbody.
Only small discrepancies remain between the two methods, which disappear linearly as the time step $\Delta t$ is decreased, method~1 generally yielding smaller errors in our simple integration scheme. 
$\Delta t$ is always made small enough so that the discrepancies remain below 1\%, which makes us confident that absolute errors due to finite-time-step errors are also at most 1\%. 

\subsection{Model validation on single-asperity contacts}
\label{sec:modelValid}
\begin{figure*}
    \centering
    \includegraphics[width=0.475\textwidth]{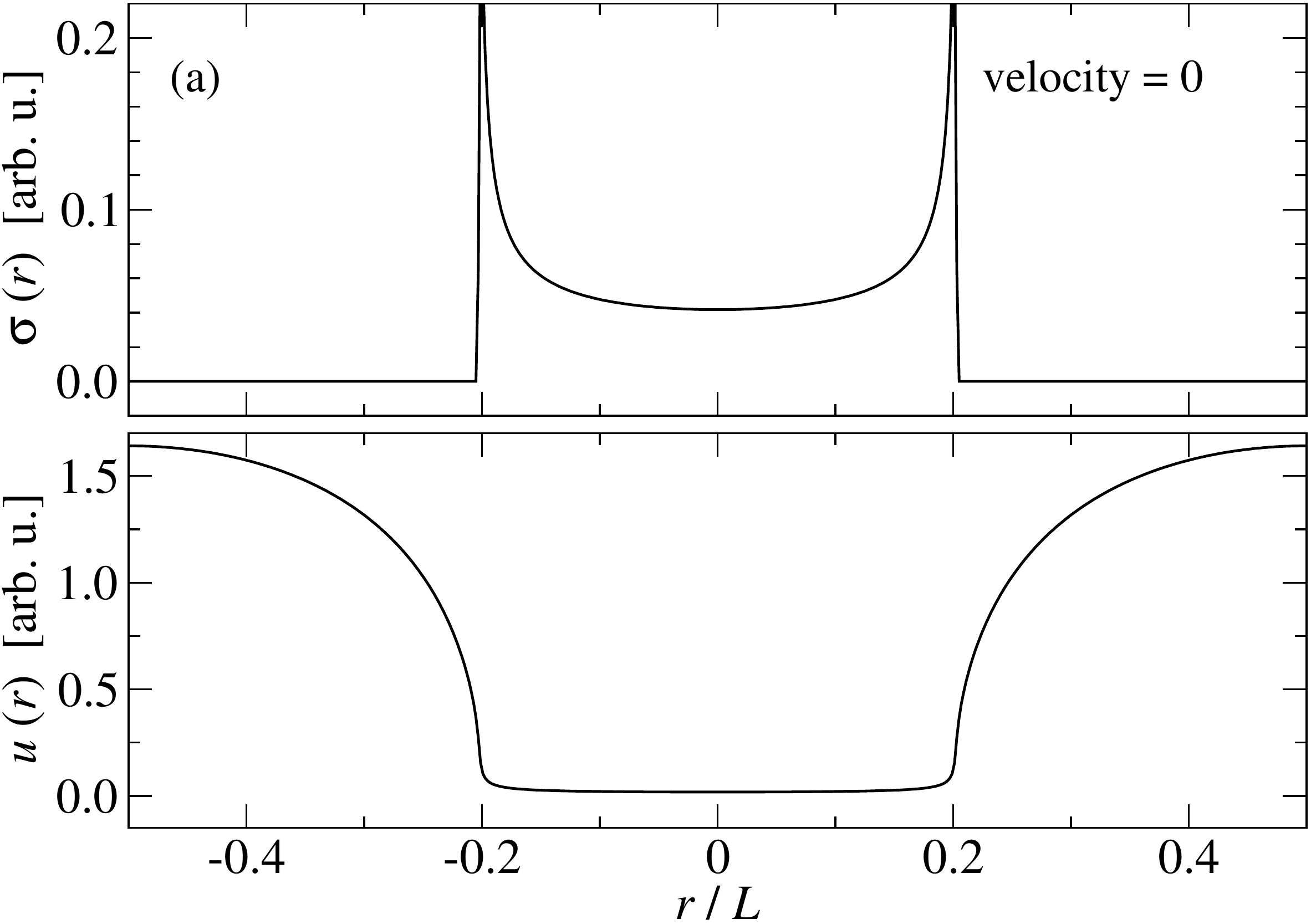}\hspace*{3mm}
    \includegraphics[width=0.475\textwidth]{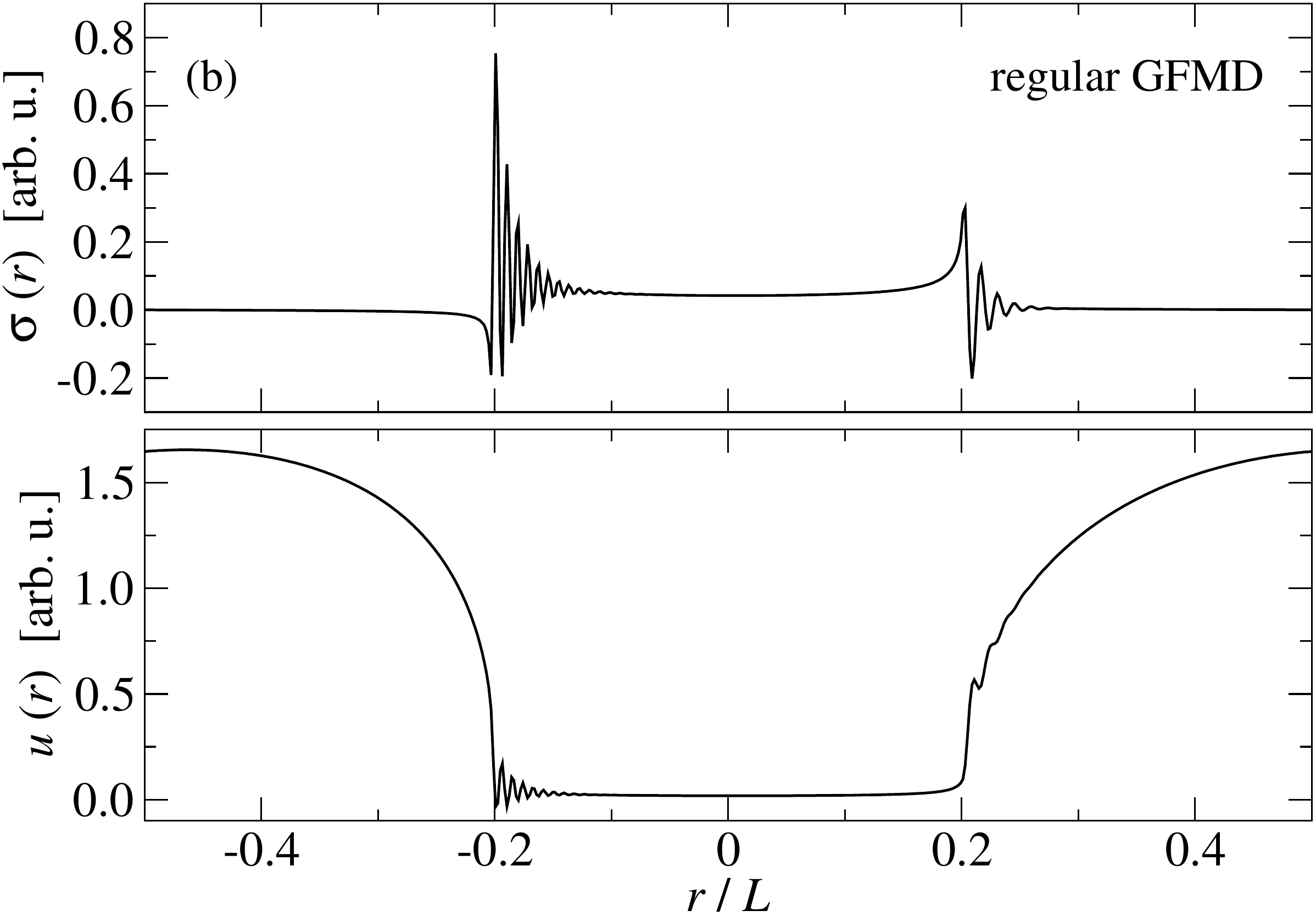} \vspace*{3mm}
    
    \includegraphics[width=0.475\textwidth]{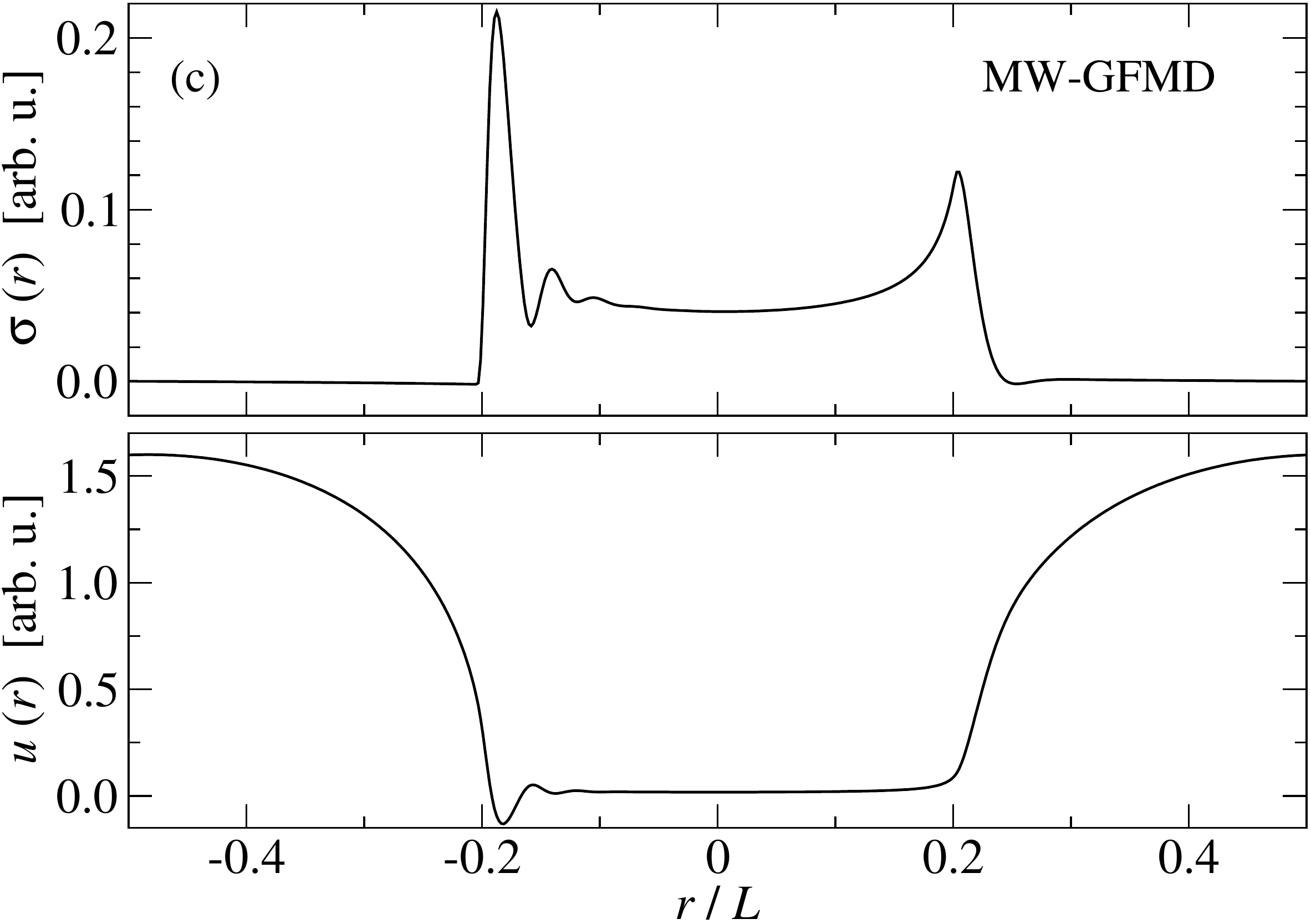} \hspace*{3mm}
    \includegraphics[width=0.475\textwidth]{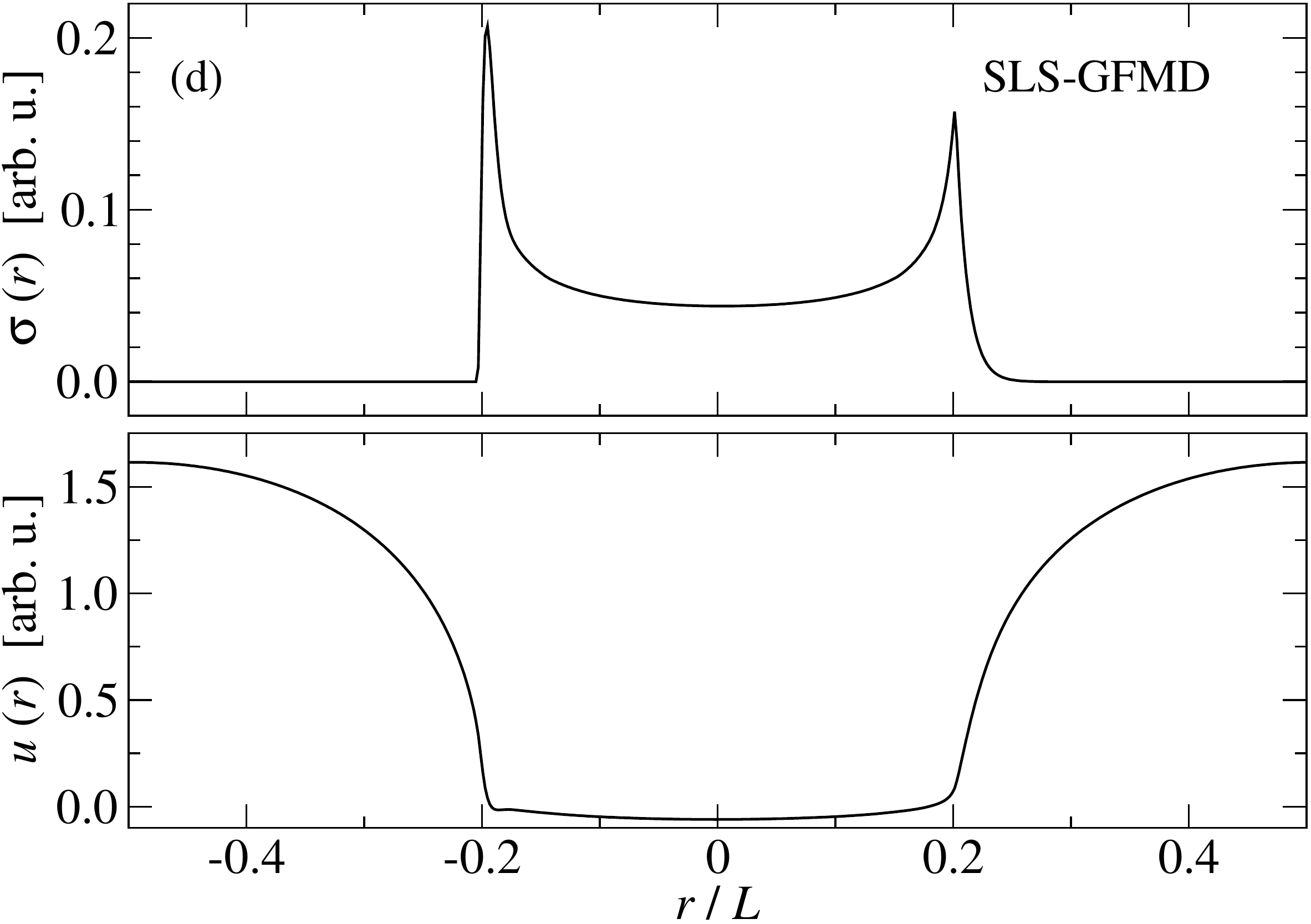}
    \caption{Stress profiles $\sigma(r)$ and displacements $u(r)$ produced by a flat-punch indenter using different dynamics for the elastomer. (a) Static equilibrium, (b) regular GFMD, (c) mass-weighting (MW) GFMD, and (d) standard-linear-solid (SLS) GFMD. The leading edge of the contact is located near $r = -L/2$, the trailing edge at $r = L/2$. The modes of the regular GFMD are (strongly) underdamped, except for the $q_0$-modes, while those of the MW-GFMD are at half the critical damping.}
    \label{fig:validation}
\end{figure*}

To validate our model and to also gain more intuition about the dynamics produced by the various viscoelastic models, we studied their time- or velocity-dependent behavior in various single-asperity contacts. 
We first discuss the sliding motion of an elastomer past a rigid, flat punch.
Its radius was set to $0.2$ times the length $L$ of the periodically repeated simulation cell.
To reduce  Gibbs ringing in the tip shape caused by the rigid translation of a discontinuous tip profile, the tip height was brought down swiftly but smoothly from its maximum value to zero. 
Results for displacements and stresses during steady-state sliding are shown in Fig.~\ref{fig:validation}

At zero sliding velocity, both stress and displacement profiles are symmetric, see Fig.~\ref{fig:validation}(a). 
The stress singularities at the edge of the contact are smoothed compared to the continuum solution due to a finite discretization and the finite stiffness of the punch-elastomer interaction. 
At non-zero velocity, the peak profiles become asymmetric for all studied dynamics, as revealed in Figs.~\ref{fig:validation}~(b--d). 
For the two dynamics involving inertia, waves are produced within the contact at the leading edge, which are much more pronounced and rugged in regular GFMD than in MW-GFMD.
Assigning larger inertia to larger wavelength undulations, i.e., by reflecting them more realistically in those cases where inertia matter, would certainly bolster this trend. 

\begin{figure*}
    \centering
    \includegraphics[width=0.75\textwidth]{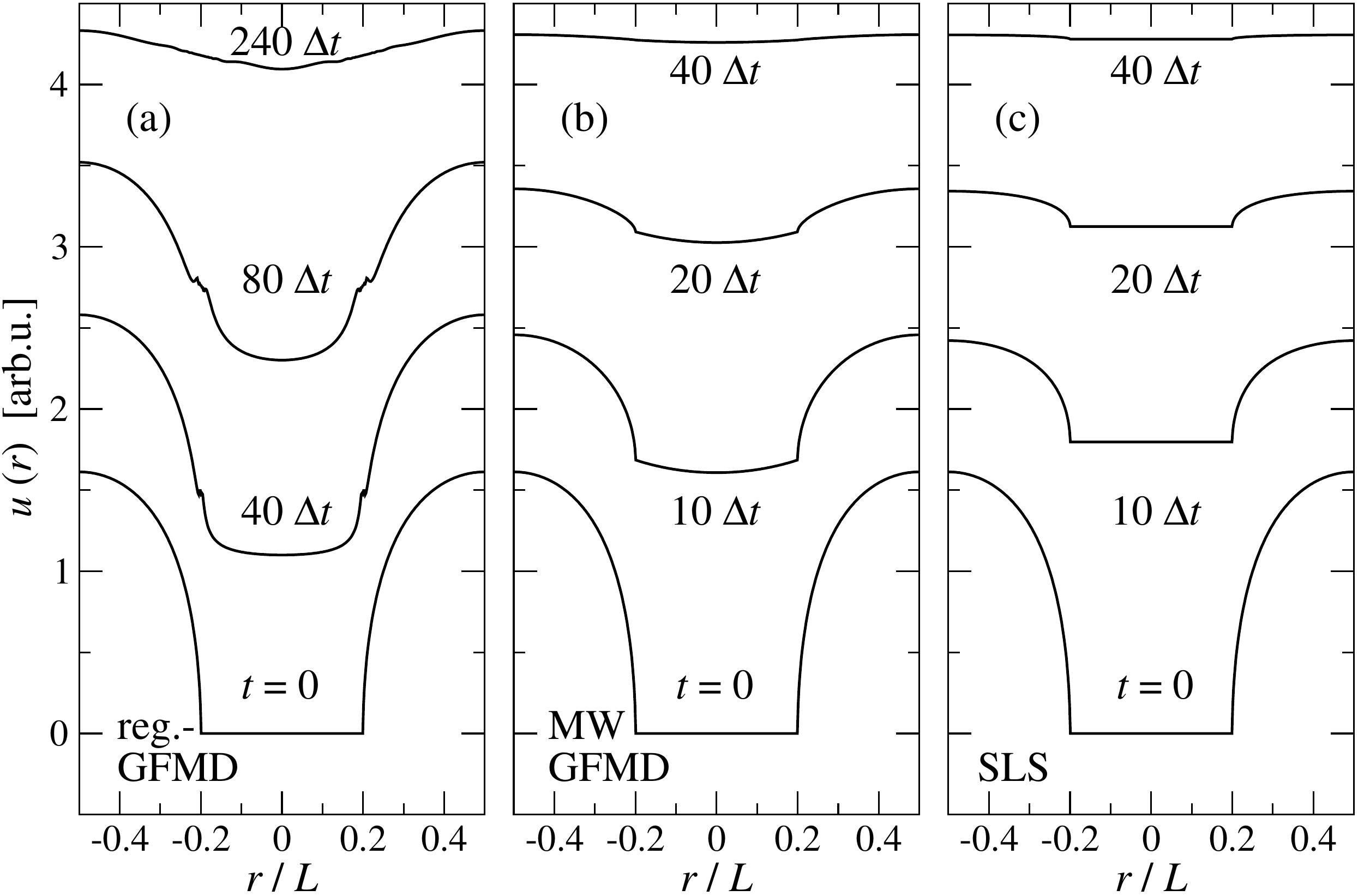}
    \caption{Relaxation dynamics from a rigid-punch indentation in the three different models, (a) regular GFMD, (b) mass-weighting (MW) GFDM, and (c) standard-linear solid (SLS) dynamics. The time evolution is shown in units of time steps, one time step corresponding to $\Delta t = 1$.}
    \label{fig:relaxRigidP}
\end{figure*}

Further insight into the dynamics of the various models can be gained from Fig.~\ref{fig:relaxRigidP}, which shows how the displacement fields relax after the sudden removal of a flat punch.
MW-GFMD and SLS dynamics look similar to the eye.
This is because a free surface has relaxation times that 
are independent of the wavevector for SLS dynamics and almost independent for MW-GFMD dynamics.
As a consequence, SLS is a perfectly shape conserving relaxation, while MW-GFMD produces only an almost shape-conserving relaxation, albeit with slightly different dynamics than SLS.
Specifically, the relaxation function is a single exponential for SLS, while it is (close-to) critical damping dynamics for MW-GFMD.
In contrast, regular GFMD makes short-wavelength undulation move on smaller time scales than long-wavelength undulations. 
The time step is therefore restricted by local dynamics so that it takes many times steps for the coarse features of the profile to disappear. 

To ascertain which dynamical model mimics what application, it must be determined when inertia and when damping counteracts the static restoring force of a surface undulation more strongly for a given wave number $q$ and velocity $v$. 
Their relative importance can be crudely estimated with the ratio 
$r(q) = \rho\, \omega / (q^2 \, E_2\, \tau) \to \rho\,v/(qE_2\, \tau)$,
so that inertial corrections have the upper hand for $r(q) \gg 1$ and viscous corrections for $r(q) \ll 1$.
Assuming typical values for moderately cross-linked elastomers, say,  $E^* = O(10$~MPa), $\tau = O(1$~s), $\rho \approx 10^3$~kg/m$^3$, and sliding velocities of $v = 10^{-3}-1$~m/s, the cross-over wave number $q_\textrm{c}$, for which $r(q_\textrm{c})\approx 1$ is many orders of magnitude, i.e., more than seven decades, smaller than typical roll-off wave numbers $q_\textrm{r} \approx O(10^4$~m$^{-1}$). 
Thus, inertial effects are clearly irrelevant for sliding dynamics of elastomers with relaxation times of order 1~s. 
However, for hard-matter systems, in which relaxation does not arise from thermally activated dynamics but from phonon-phonon or phonon-electron coupling, $\tau$ is easily nine to twelve orders of magnitude smaller than for polymers.
Inertia would then prevail even at sub-micrometer scales. 
However, other effects, such as plastic deformation, may have to be included in order to faithfully represent the time dependence of contact stresses locally.
At large wavelengths, these details may be negligible owing to Saint-Venant's principle. 

\begin{figure}
    \centering
    \includegraphics[width=0.475\textwidth]{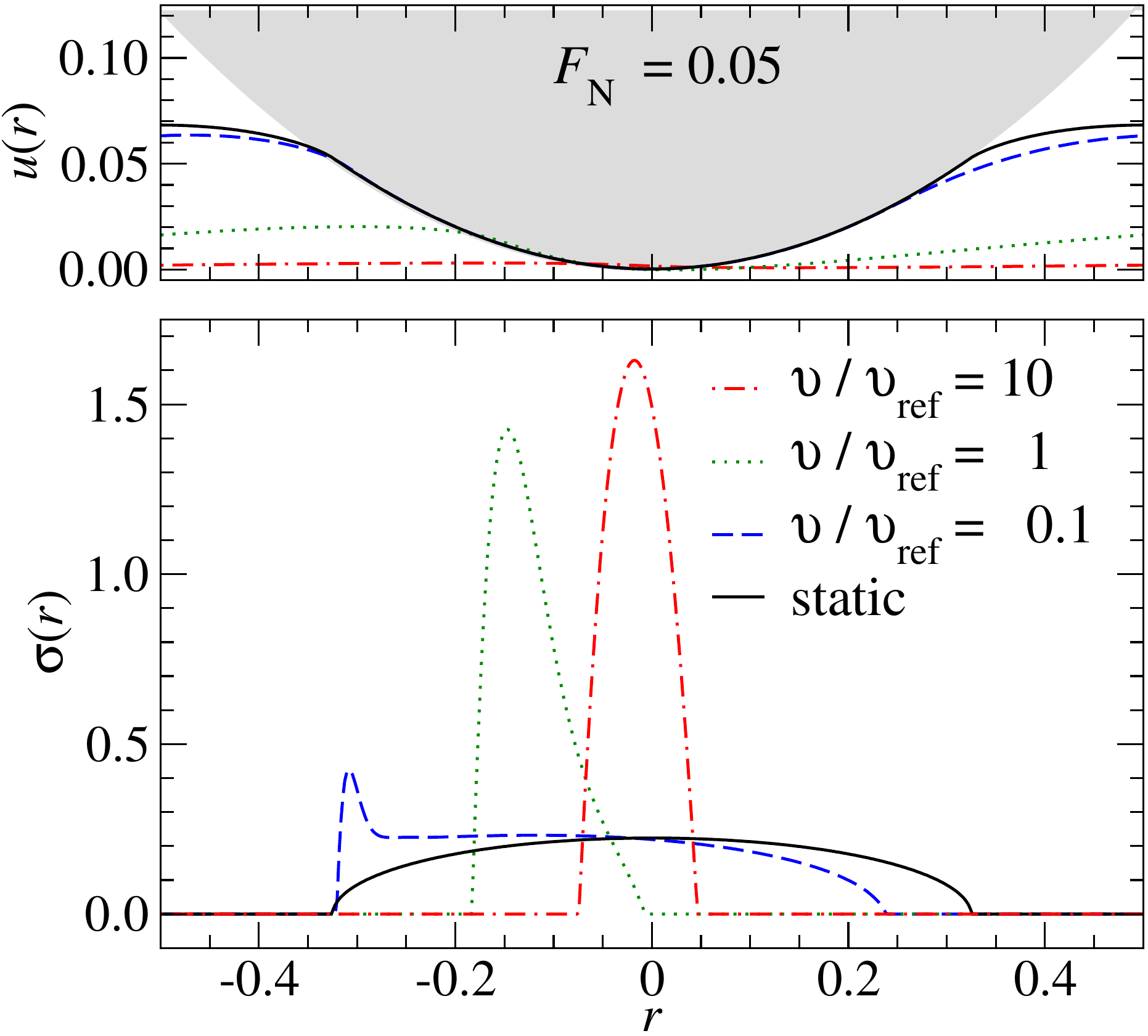}
    \caption{\label{fig:asymmHertz}
     Contact geometry (\textbf{top}) and  interfacial stress (\textbf{bottom}) during steady-state sliding of the standard-linear-solid model with $E_1/E_2 = 1000$ past a Hertzian indenter at different sliding velocities. 
    Units are defined such that the static contact modulus $E^*$, the relaxation time $\tau$, and the radius of curvature $R_\textrm{c}$ are all unity.
    }
\end{figure}

In the final part of the model validation, a Kelvin-Voigt elastomer slides past a Hertzian tip at different velocities.
At low and high velocities the contact is close to being circular in shape, and the surface stresses resemble that of a Hertzian contact (see Fig.~\ref{fig:asymmHertz}).
This could be rationalized in a straightforward way when considering the two limiting cases of frequency-dependent elastic modulus of the viscoelastic modulus: at very low speed it is almost constant and is close to $E_2$, while at high velocities the modulus effectively approaches close to $E_1$.
These observations are perfectly in line with results of earlier work~\cite{Panek1980JAM,Carbone2013JMPS,Scaraggi2014TL}.

\section{Theory}
\label{sec:theory}

\subsection{Full contact}
In our model, full contact is achieved when the pressure is high enough to induce a negative gap throughout the contact. 
The Fourier component of the stress acting on the surface is
\begin{equation}
    \tilde{\sigma}(\mathbf{q},t) = \kappa_\textrm{i} \,
    \left\{ \tilde{h}(\mathbf{q},t) - \tilde{u}(\mathbf{q},t)  \right\} + p_0\,\delta_{\mathbf{q},0},
\end{equation}
so that the solution for the displacement $\tilde{u}(\mathbf{q},t)$ in the frequency domain satisfies
\begin{equation}
    \tilde{u}_\textrm{fc}(\mathbf{q},\omega) = 
    \frac{\kappa_\textrm{i}}{\kappa(q,\omega) + \kappa_\textrm{i}} 
    \, \tilde{h}(\mathbf{q},\omega),
    \label{eq:solutionMode}
\end{equation}
for $\mathbf{q}\ne 0$, where $\kappa(q,\omega)$ is the effective ``stiffness'', which satisfies
\begin{equation}
    \kappa(q,\omega) = 
    \begin{cases}
    \frac{qE^*}{2} - m_q\omega^2 +
    i\gamma m_q \omega & \textrm{(conv. GFMD)} \\
\frac{q}{2}  \,
\frac{(1+i\omega\tau)\,E_2 \,E_1}{(1+i\omega\tau)E_2 +E_1}& \textrm{(SLS).}
    \end{cases}
\end{equation}
The regular full-contact solution,
$\tilde{u}_\textrm{fc}(\mathbf{q},\omega)=\tilde{h}(\mathbf{q},\omega)$,
is recuperated in Eq.~(\ref{eq:solutionMode}) in the limit of infinitely large local interfacial stiffness, i.e. when $\kappa_\textrm{i} \to \infty$.

At constant $\mathbf{v}$, the frequency associated with a wave vector $\mathbf{q}$ is $\omega = \mathbf{q}\cdot\mathbf{v}$ so that the absolute square of the stress acting on that mode at constant sliding velocity $\textbf{v}$ is
\begin{equation}
    \left\langle  \left\vert \tilde{\sigma}_\textrm{fc}(\mathbf{q}) 
    \right\vert^2 \right\rangle_\mathbf{v} = 
    \left\vert  \frac{\kappa(q,\mathbf{q}\cdot\mathbf{v}) \,\kappa_\textrm{i}}
    {\kappa(q,\mathbf{q}\cdot\mathbf{v}) + \kappa_\textrm{i}} \right\vert^2 \,
    \left\vert \tilde{h}(\mathbf{q}) \right\vert^2.
    \label{eq:stressBroad}
\end{equation}
This expression, which will be needed later, states by how much the second moment of the stress distribution increases in full contact due to the existence of the height Fourier component $\tilde{h}(\mathbf{q})$.

The full-contact dissipated power remains to be determined.
Towards this end, it is useful to realize that each mode contributes with
\begin{eqnarray}
P_\textrm{d}^\textrm{fc}(\mathbf{q},\mathbf{v}) = A\,
\kappa''(q,\mathbf{q}\cdot\mathbf{v})\,
(\mathbf{q}\cdot\mathbf{v})^2\,
\left\vert \tilde{h}(\mathbf{q}) \right\vert^2 \,
\end{eqnarray}
to the total dissipated power, where $\kappa''(q,\omega)$ is the imaginary part of $\kappa(q,\omega)$. 

\subsection{Partial contact}
In Persson's theory, a contact problem is first solved without considering random roughness. 
Thus, for flat, periodically repeated surfaces, the stress distribution is initially represented by a delta function, $\Pr(\sigma) = \delta(\sigma-p_0)$.
The next assumption is that the stress distribution would broaden in the contact by an amount given by Eq.~(\ref{eq:stressBroad})
if the roughness undulations of the nominally flat indenter associated with $h(\mathbf{q})$ were resolved and included into the calculation.
If all roughness with wave vectors of magnitude less or equal $q$ are resolved, the following interfacial stress or pressure variance is obtained
\begin{equation}
    \Delta p^2(q) = \sum_{\mathbf{q}',q'\le q} \left\langle  \left\vert \tilde{\sigma}(\mathbf{q}) 
    \right\vert^2 \right\rangle.
\end{equation}
Since negative stresses are forbidden in repulsive contacts, the stress distribution in contact, i.e., for $\sigma>0$ is estimated by
\begin{eqnarray} \label{eq:stressDistRW}
    \Pr(\sigma>0,q) & = & \frac{1}{\sqrt{2\pi\Delta p^2(q)}}
\left\{
    \exp{\left[-\frac{(p-p_0)^2}{2\,\Delta p^2(q) }\right]} \right. \nonumber \\
    & - & \left.\exp{\left[-\frac{(p+p_0)^2}{2\,\Delta p^2(q) }\right]}
   \right\}
   .
\end{eqnarray}
This distribution is motivated by the interpretation of the stress in real space as a random walk, and thus as a diffusive process, in which the stress in a contact point increases or decreases randomly whenever an additional $\tilde{h}(\mathbf{q})$ is resolved.
In this analogy, a stress of zero is an absorbing barrier in the diffusive process, where the random walker moves out of contact.
This constitutes an absorbing barrier in the diffusive process, whose effect is reflected by the second summand on the r.h.s. of Eq.~(\ref{eq:stressDistRW}).
The latter could also be called a mirror Gaussian by the mathematical analogy of the diffusion equation to electrostatic problems, where absorbing or reflecting boundaries can be represented through mirror charges. 
For a more detailed analysis of this analogy, the reader is referred to the original literature~\cite{Persson2001JCP,Persson2006SSR} or to a derivation similar in spirit to the one presented here~\cite{Dapp2014JPCM}.

The relative contact area, which would be obtained if only those height undulations with wave vectors whose magnitude does not exceed $q$ were resolved, can then be obtained by an integral over the stress distribution function over positive $\sigma$ so that
\begin{equation} \label{eq:relCont}
    a_\textrm{r}(p_0,q,\mathbf{v}) = \mathrm{erf}
    \left( \frac{p_0}{\sqrt{2}\,\Delta p(q,\mathbf{v})}\right).
\end{equation}

If we now assume that forced motion and thus dissipation occurs predominantly in the contact area, the magnitude of the friction force becomes
\begin{equation}
\label{eq:frictionPartial}
f = \frac{1}{v\,}
\sum_{\mathbf{q}}
W\left[ a_\textrm{r}(p_0,q,\mathbf{v})\right]\,
P_\textrm{d}^\textrm{fc}(\mathbf{q},\mathbf{v}),
\end{equation}
where the weight function $W(a_\textrm{r})$ was originally chosen to be $W(a_\textrm{r}) = a_\textrm{r}$, while later modifications of the theory assumed $W(a_\textrm{r}) = \gamma\, a_\textrm{r} + (1-\gamma)\,a_\textrm{r}^3$ with $\gamma$ being a numerical constant~\cite{Persson2006SSR}.
Reasons for why the weight function $W(a_\textrm{r})$ might be less than the relative contact area---and not simply coincide with it---were proposed recently~\cite{Muser2021TL}.

\section{Results}
\label{sec:results}

Before presenting numerical results and comparing them to Persson's theory, we need to emphasize that two dimensionless numbers of order unity were used in the theory, which, however, remained constant throughout all calculations. 
First, for the weighting function $W(a_\textrm{r})$ introduced in Eq.~(\ref{eq:frictionPartial}), $\gamma = 0.6$ is used rather than 0.4, which is commonly assumed to accurately predict the elastic energy of semi-infinite elastomers~\cite{Persson2007PRL}.
The reason for the need of such a correction factor has recently been linked to the observation that the root-mean-square gradient averaged over true contact is less than its average over the entire domain when true contact is partial~\cite{Muser2021TL}.
Since the theory is not an exact theory, except in full contact, the optimum numerical value for $\gamma$ may thus differ for the calculation of elastic and kinetic energy. 
Second, to better match results on relative contact area, the prefactor $\alpha = 1.25$ precedes the nominal pressure $p_0$ when computing the relative contact area with Eq.~(\ref{eq:relCont}), i.e., we use $a_\textrm{r}(\alpha\, p_0, q, \mathbf{v})$ when computing $W(a_\textrm{r})$.
This corrections makes the predicted low-pressure relative contact area obey $a_\textrm{r} \approx 2\, {p}_0^*$ at small reduced pressure $p_0^* \equiv \,p_0/(\bar{g} E^*) \lesssim 0.1$, where $\bar{g}$ is the root-mean-square gradient of the rigid indenter---at least when both ratios $\lambda_\textrm{s}/\lambda_\textrm{r}$ and $\lambda_\textrm{r}/L$ are very small~\cite{Wang2021TL}. 

\begin{figure*}
    \centering
    \includegraphics[width=0.45\textwidth]{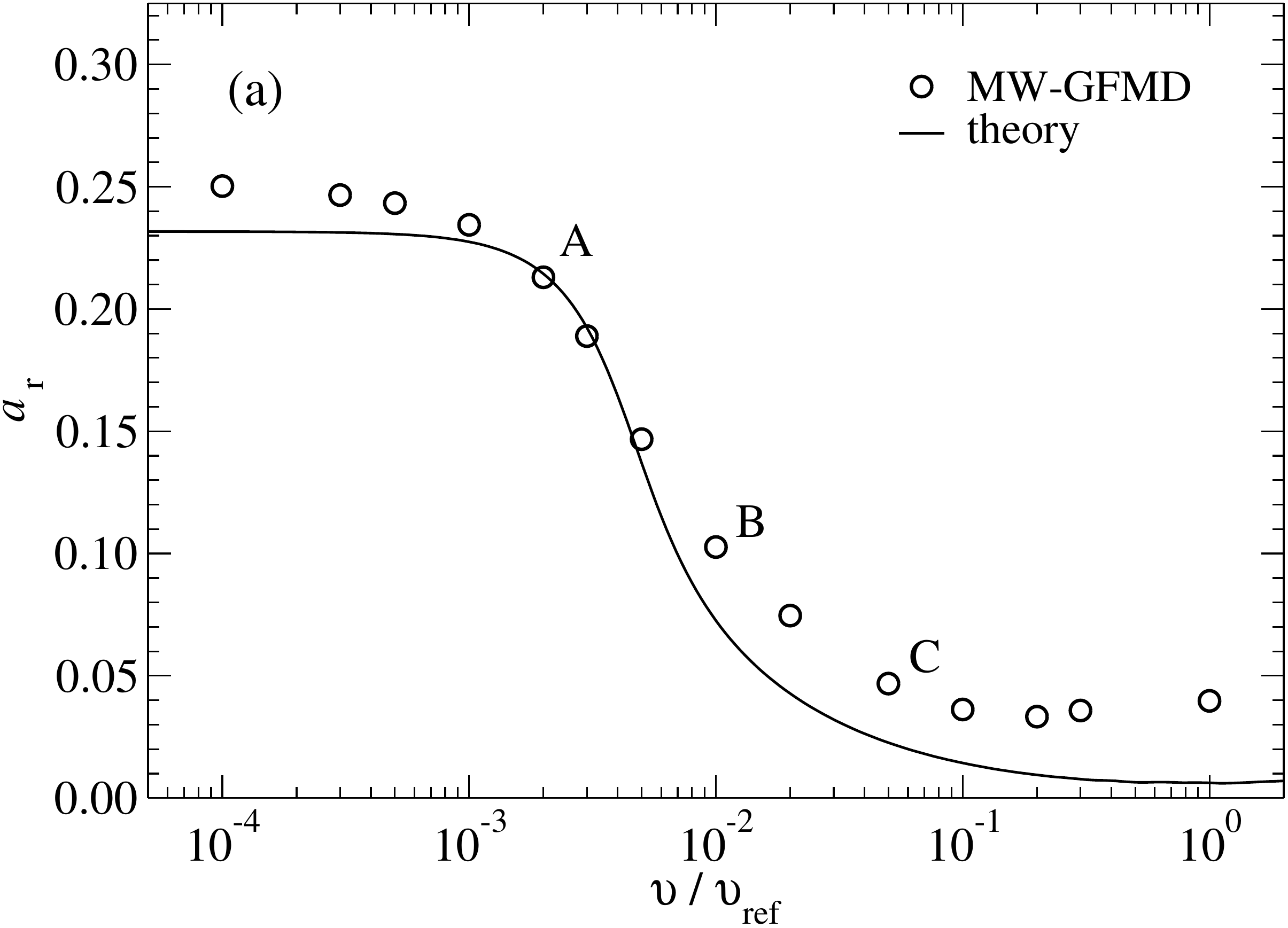}
    \includegraphics[width=0.45\textwidth]{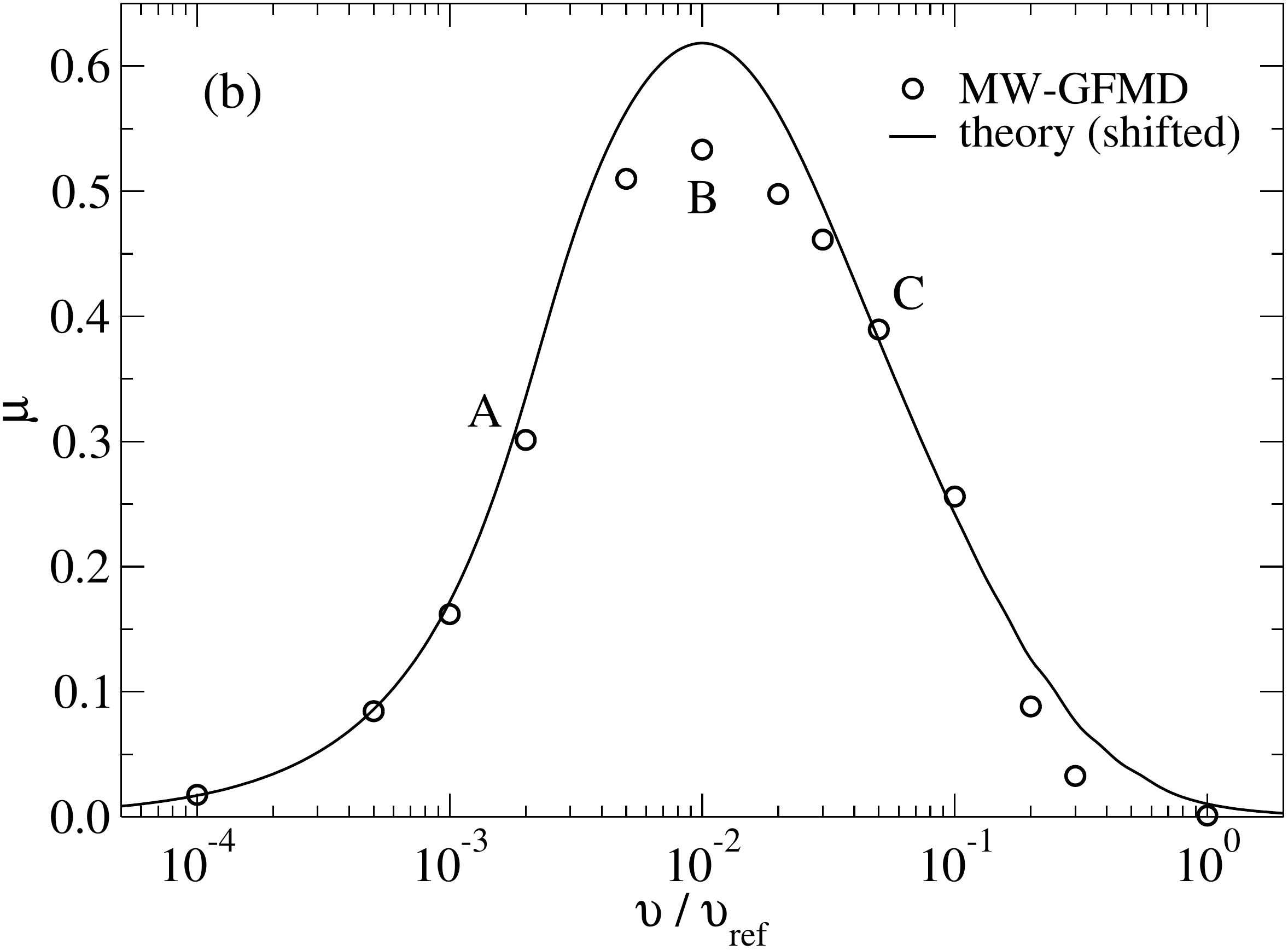}
    \includegraphics[width=0.32\textwidth]{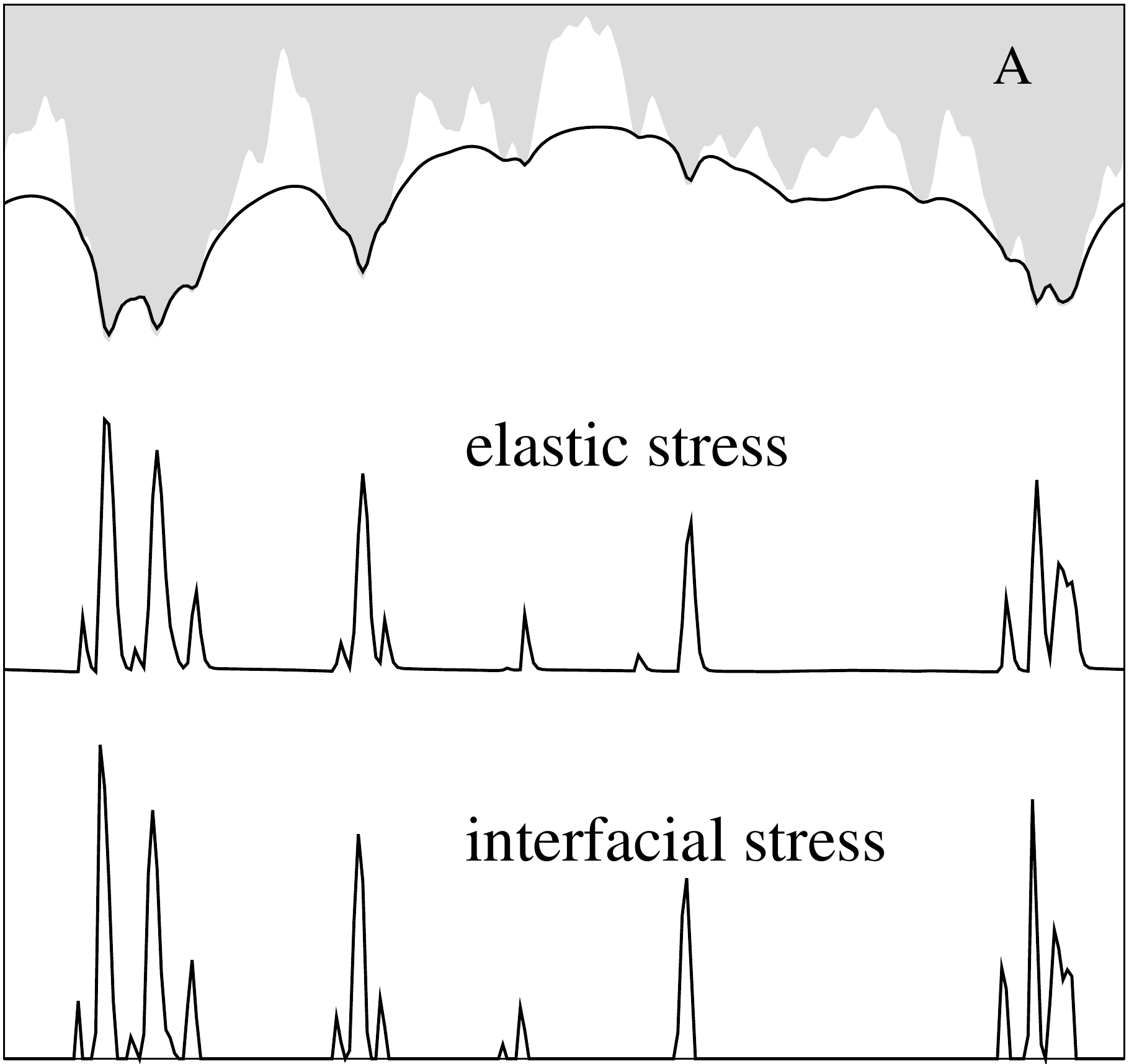}
    \includegraphics[width=0.32\textwidth]{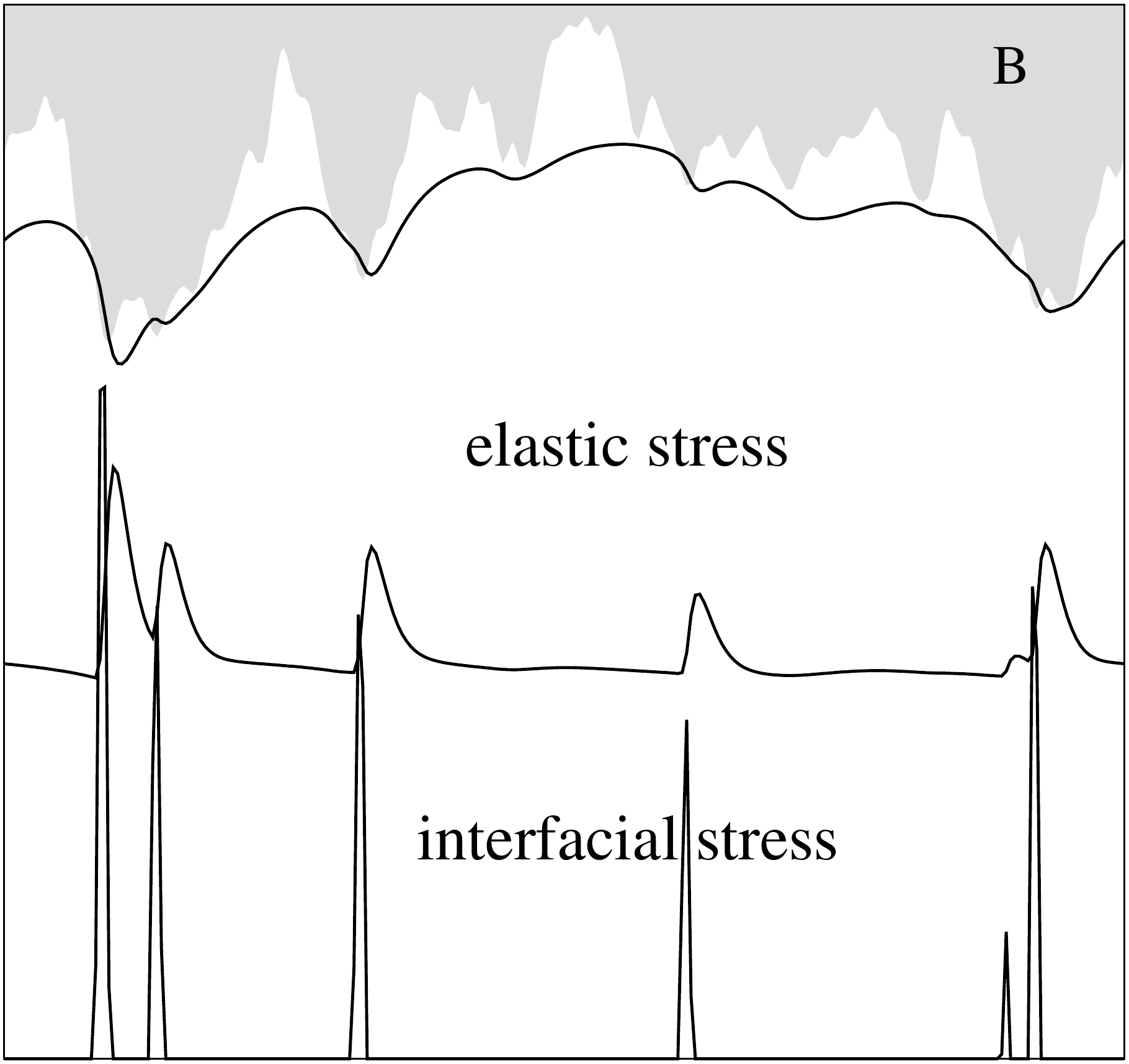}
    \includegraphics[width=0.32\textwidth]{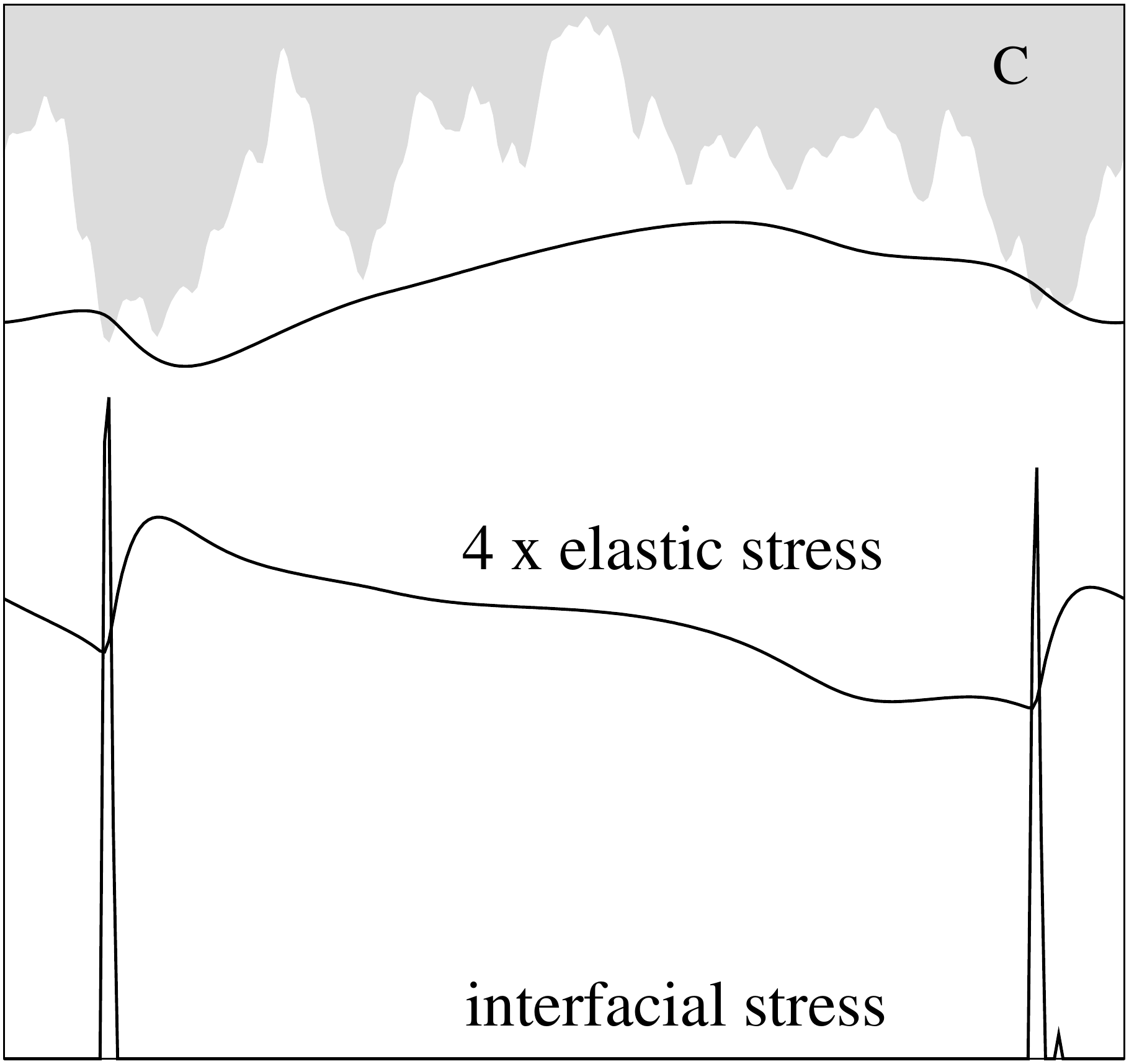}
    \caption{\textbf{Top row:} (a) Relative contact area $a_\textrm{r}$ (\textbf{left}) and (b) friction force $F$ (\textbf{right}) as a function of velocity $v$. The theoretical prediction for the force is shifted to the left by plotting it as $F(v/2)$ to demonstrate qualitative similarity between theory and simulation. 
    \textbf{Bottom row:} Contact geometry and normal elastic as well interfacial stresses at low (A), intermediate (B), and large (C) sliding velocity. 
    Letters used in the bottom row correlate with those in the top row.
    The size of the simulation cell is reduced compared to that of the default model. The surface modes of a free surface are set to be critically damped, i.e., $\gamma = 2$.}
    \label{fig:MWP0_1S}
\end{figure*}

To set the stage for further discussion, we first juxtapose theoretical and numerical results for a small system in Fig.~\ref{fig:MWP0_1S}. 
Its size and dynamics are altered with respect to the default model for illustrative purposes.  
Specifically, the ratio $q_\textrm{s}/q_\textrm{r}$ is reduced by a factor of four compared to the default system, allowing the whole system to be visualized while still resolving small-scale features.
Moreover, the dynamics are those obtained by mass weighting, which makes it possible to reveal inertial effects.
Yet, many of the observations that can be made on the investigated small, inertial system repeat themselves for larger systems with conventional viscoelastic dynamics.

First, it can be noticed in panel (a) of Fig.~\ref{fig:MWP0_1S}
that the theory reproduces the relative contact area as a function of velocity reasonably well up to intermediate velocities, where the contact area decreases due to the sliding motion.
The velocity at which $a_\textrm{r}$ is reduced to half its static value is well matched.
However, the theory clearly underestimates the relative contact area at very large velocities. 
At the same time, albeit, barely visible to the eye, a relative minimum occurs in $a_\textrm{r}$ at large $v$.
This effect is somewhat more ascertainable in the numerical data than in the theory. 

Next, it can be seen in the bottom row of Fig.~\ref{fig:MWP0_1S} that differences between elastic and interfacial stress are minor at small velocities.
They become larger with increasing sliding velocity.
At very large velocities, the (vertical) velocity of a surface element, which is lateral velocity times the gradient of the steady-state displacement field, changes discontinuously after being impacted by a bump on the rough surface. 
Due to inertia, the maximum displacement caused by the momentum transfer does not peak right at the end of the impact, as it does for standard dynamics, but shortly after. 
In either case, the elastic stress does not relax quasi-instantaneously so that the dissipation caused by the relaxation occurs at points of time when the impacted point has moved far away from the asperity causing the impact. 
As a consequence, at very large velocities, elastic and interfacial stress no longer resemble each other. 
In our understanding, this retardation effect is not accounted for in Persson's theory, which, in our opinion, is the main reason why the relative differences between numerical and theoretical friction coefficients are large as revealed in panel (b) of Fig.~\ref{fig:MWP0_1S}. 
At the same time, one may wonder if this deficiency in the theory needs to be fixed as absolute errors are small.

\begin{figure*}
    \centering
    \includegraphics[width=0.475\textwidth]{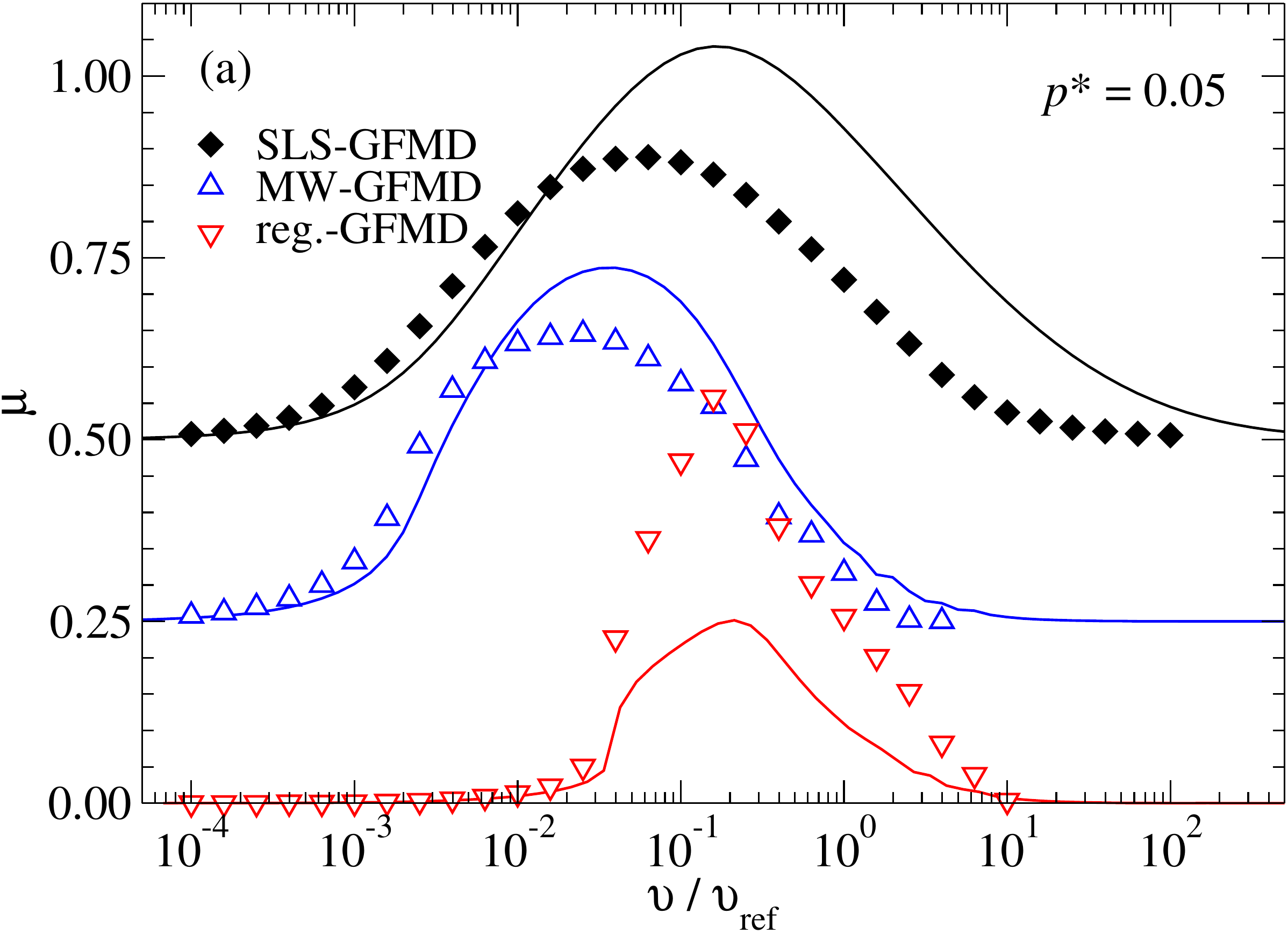}
    \includegraphics[width=0.475\textwidth]{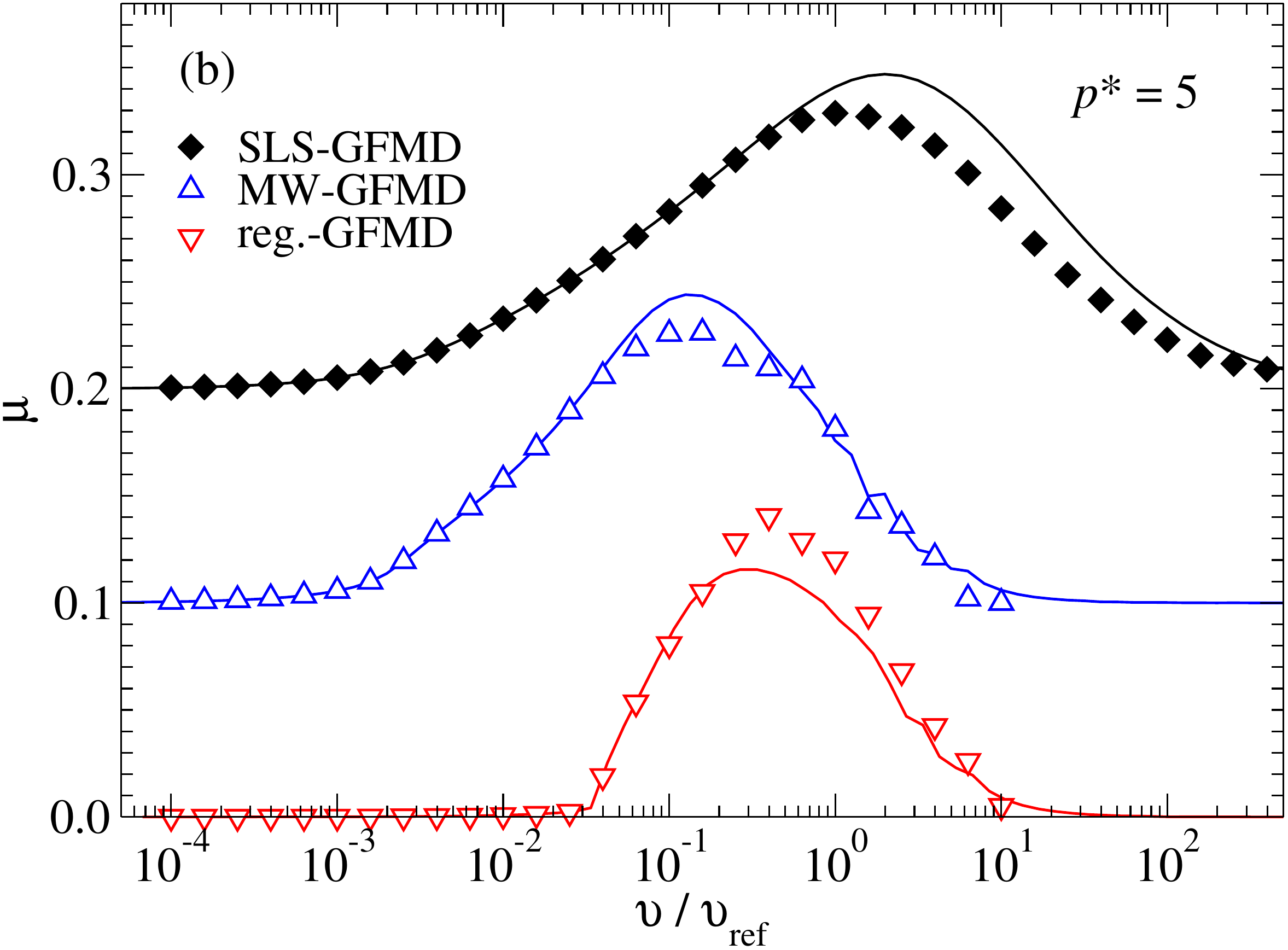}
    \caption{\label{fig:dynamicalModels}
    Friction coefficient $\mu$ as a function of velocity for different dynamical models: regular GFMD (triangle down, red), mass-weighted-GFMD (triangle up, blue), and standard-linear solid (SLS) dynamics (filled diamond, black). Velocity is expressed in units of $v_{\rm ref} = \gamma\,\lambda_\textrm{r}$ for regular and MW-GFMD dynamics and in $v_{\rm ref} = \lambda_\textrm{r}/\tau$ for SLS dynamics.
    MW and SLS curves are shifted to avoid overlap between lines. Both tend to zero in the limit of zero velocity. 
    \textbf{Left:} Low-pressure, $p^* \equiv p/(\bar{g}E^*)  = 0.05$, leading to roughly 10\% relative contact area. Symbols represent GFMD results, lines - theory. Slider geometry and substrate-slider interactions are fixed to their default values.
    \textbf{Right:} High-pressure, $p^* = 5$, leading to full contact at zero sliding velocity.}
\end{figure*}

\begin{figure*}
    \centering
    \includegraphics[width=0.475\textwidth]{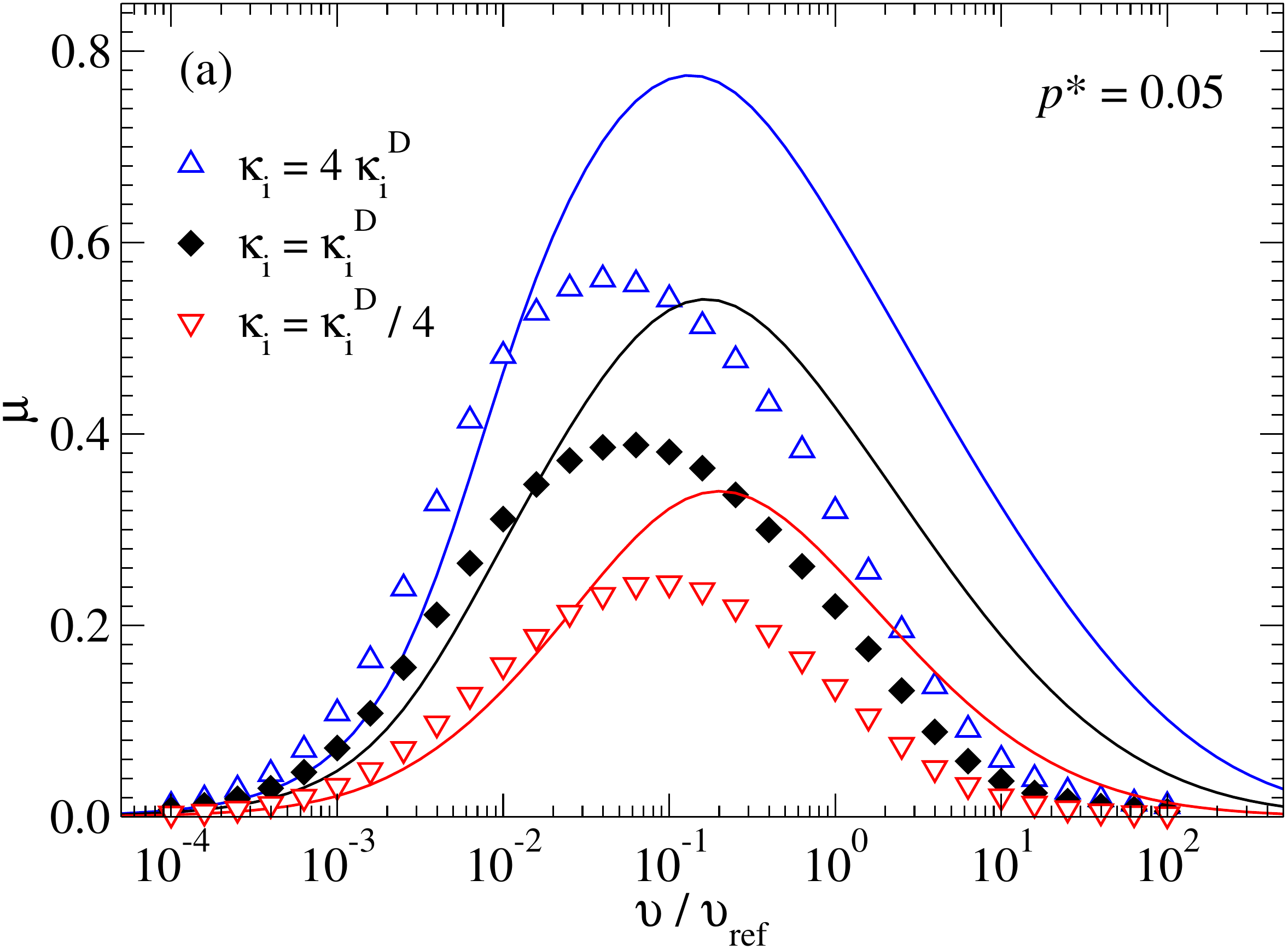}
    \includegraphics[width=0.475\textwidth]{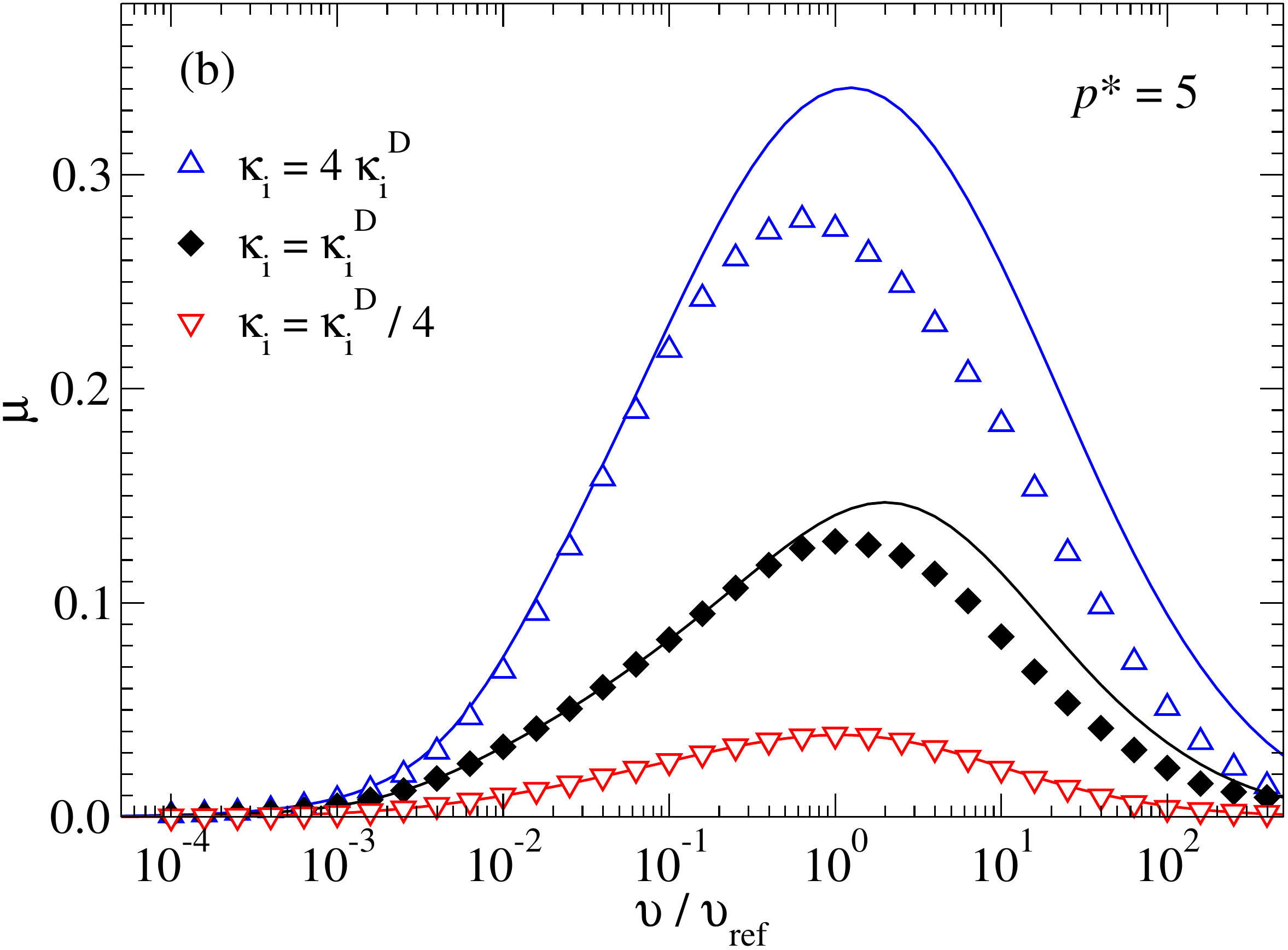}
    \caption{\label{fig:interactionModels}
    Similar to Fig.~\ref{fig:dynamicalModels}, however, this time, the dynamical model is fixed to SLS dynamics and the substrate-slider interactions are varied.
    In addition to the default value of $\kappa_\textrm{i}^{\rm D} = 0.2\,q_\textrm{max}\,E_2$ (filled diamond, black), the numerical values
    $0.25\kappa_\textrm{i}^{\rm D}$ (triangle down, red) and $4\kappa_\textrm{i}^{\rm D}$ (triangle up, blue) are considered.
    }
\end{figure*}

\begin{figure*}
    \centering
    \includegraphics[width=0.475\textwidth]{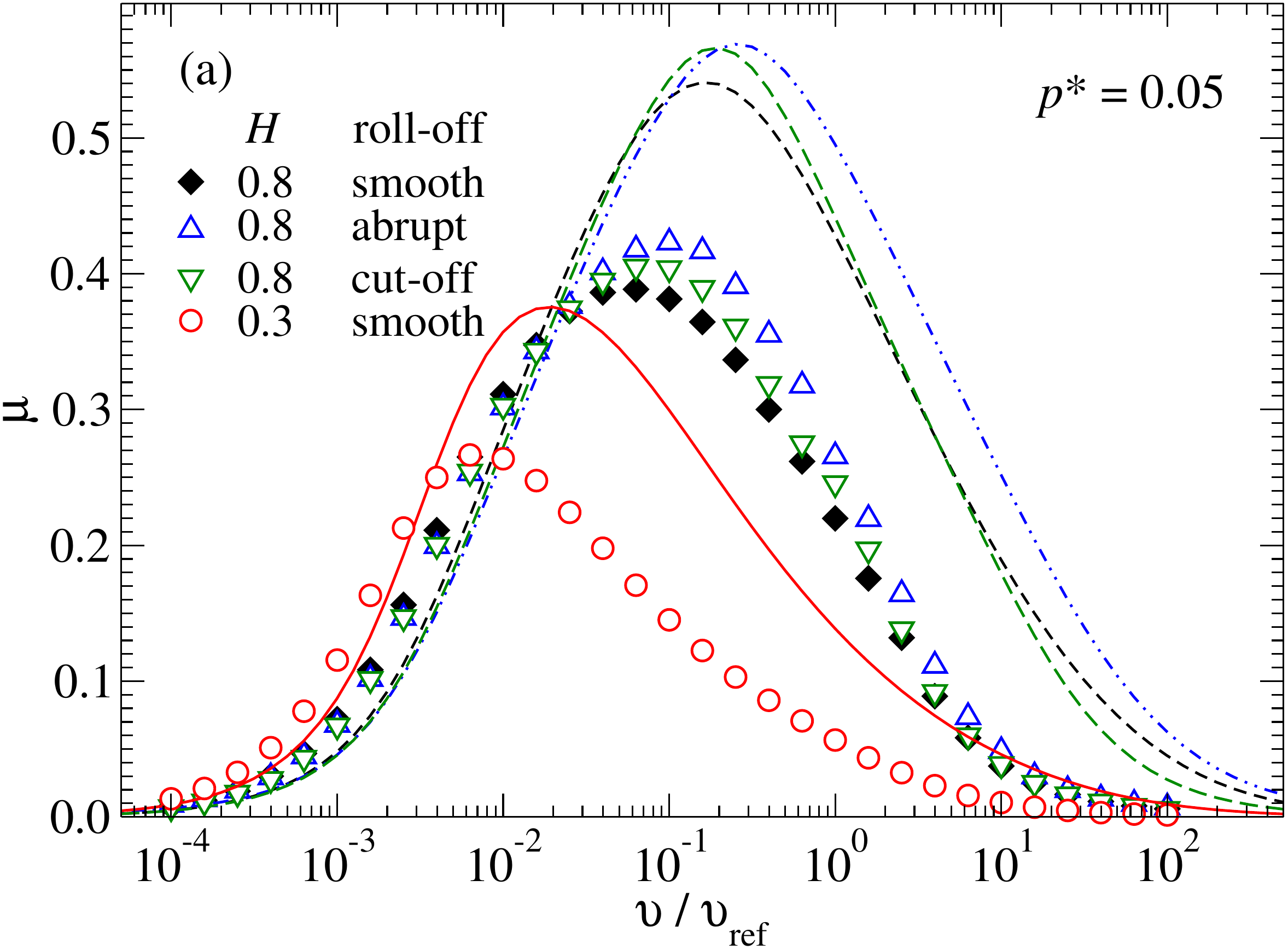}
    \includegraphics[width=0.475\textwidth]{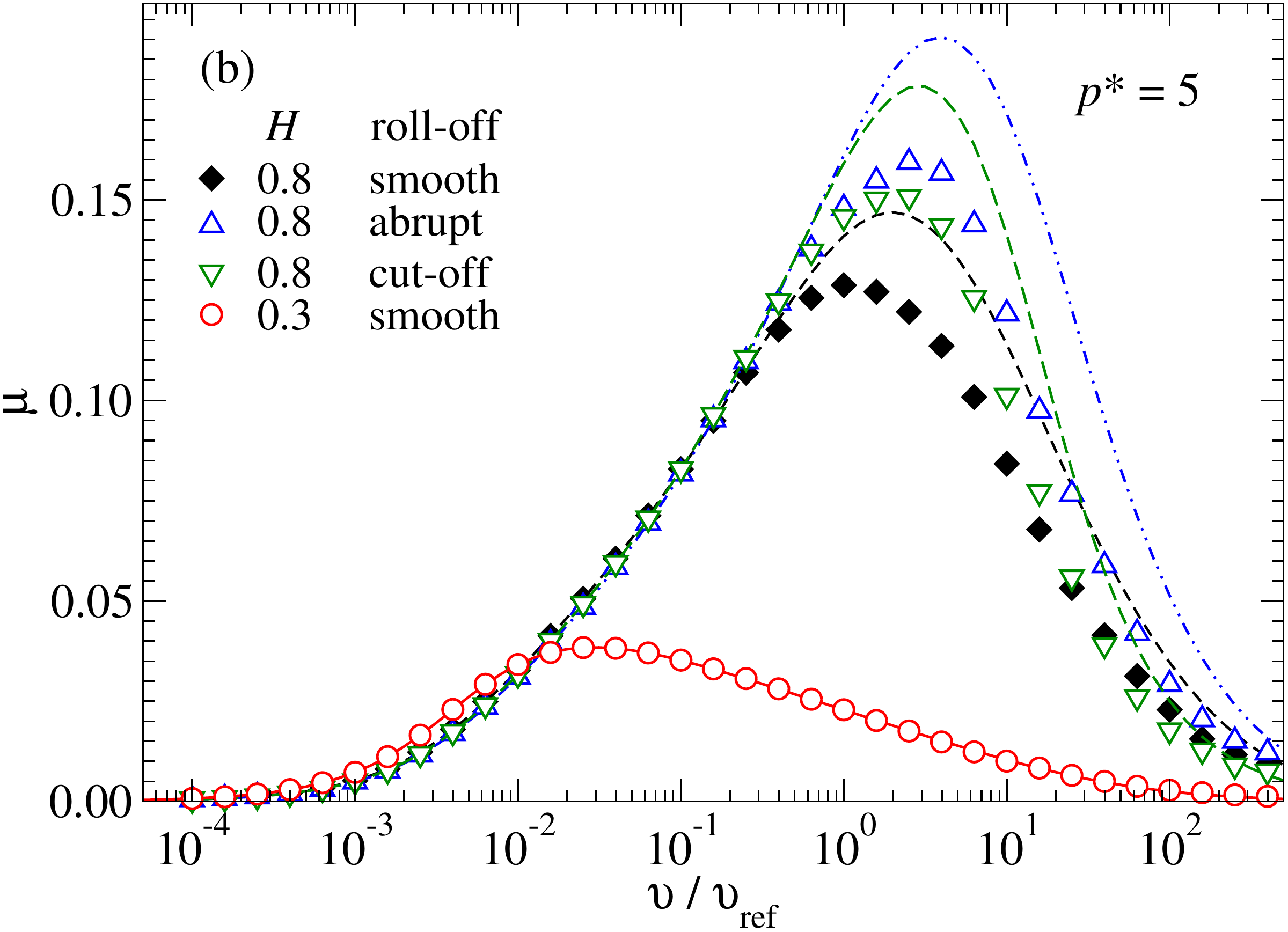}
    \caption{\label{fig:substrateModels}
     As Fig.~\ref{fig:dynamicalModels}, however, this time, the dynamical  model is fixed to SLS dynamics and the substrate geometry is varied. Friction as a function of velocity for $H=0.8$ with smooth roll-off (triangle up, red), $H=0.8$ with regular roll-off (triangle down, blue), $H=0.8$ with cut-off (diamond, green) and $H=0.3$ with smooth roll-off (circle, black). 
     $\lambda_\textrm{r}/L$, 
     $\kappa_\mathrm{i}$, $\Delta a/\lambda_\textrm{max}$ were fixed to their default values.
     Velocity is expressed in units of $v_{\rm ref} = \lambda_{\textrm{r}}/\tau$.
    }
\end{figure*}

It may also be worth noting that the symmetry of the interfacial stress revealed by the furthest most left asperity in contact changes from almost symmetric, at least after blurring the individual stress peaks, to clearly asymmetric, and back to being almost symmetric.
This is qualitatively similar to the situation described for Hertzian contacts in Fig.~\ref{fig:asymmHertz}.

In the following, we wish analyze how well different aspects of the studied models are reflected by Persson's theory.
To this end, we will include a default set-up and vary either (i) the viscoelastic properties of the elastomer (Fig.~\ref{fig:dynamicalModels}) or (ii) the slider-indenter interactions (Fig.~\ref{fig:interactionModels}) or (iii) the substrate geometry (Fig.~\ref{fig:substrateModels}).
The default system is defined as follows:
(i) Viscoelastic properties: SLS dynamics with 
$E_1/E_2 = 10^3$ and $\tau = 1$.
The static contact modulus $E^* = E_2 E_1/(E_2+E_1)$ is set to unity. 
(ii) Slider-indenter interactions: The stiffness of the slider-indenter overlap-penalty potential is set to $0.2\,E^* q_\textrm{max}$, where $q_\textrm{max} \equiv \sqrt{8}\pi/\Delta a$ is the largest wave number of the discrete elastic manifold.
The default value for $q_\textrm{max}$ is $\gtrsim 8\pi/\lambda_\textrm{s}$ so that the mesh size satisfies $\Delta a \lesssim \lambda_\textrm{s}/4$.
This discretization is certainly not small enough to make the calculations approach the true continuum limit, which, however, we see as unproblematic for mainly three reasons.
First, nature is not continuous at the smallest scale either.
Second, we adopt the theory to account for finite contact or overlap stiffness.
Third, the theory is not an exact theory and meant to predict trends.
(iii) Substrate geometry:
Self-affine roughness with a Hurst exponent of $H = 0.8$ and a smooth roll-off.
The ratio of roll-off wavelength and system size and that of
the short wavelength cutoff and the roll-off wavelength are
$\lambda_\textrm{r}/L = 0.4$ and
$\lambda_\textrm{s}/\lambda_\textrm{r} = 1/100$, respectively.

We begin the comparison between GFMD simulations and Persson's theory  in Fig.~\ref{fig:dynamicalModels} by analyzing how different viscoelastic models affect the dependence of the friction coefficient $\mu \equiv F/L$ on sliding velocity $v$.
In addition to the default SLS dynamics, we study regular GFMD dynamics as well as mass-weighted (MW) GFMD.
For both additional types of GFMD simulations, the damping constant and the reference mass were set to $\gamma = 1$ and $m_\textrm{ref} = q_\textrm{max}\,E^*/2\,[t]^2$, which can be associated with the inertia of short wavelength modes.
To remind the reader, we mention that inertia of long wavelength modes are decreased in MW-GFMD, in order to make different modes relax on similar time scales. 

Fig.~\ref{fig:dynamicalModels} reveals the generic behavior of rubber friction for all dynamical models and at both high and low pressure:
$\mu$ takes its maximum at intermediate $v$.
Theory and simulations correlate quite well.
Agreement is almost perfect at large reduced pressures and small sliding velocity.
This is not surprising, as the theory uses the full-contact solution as input.
The agreement remains semi-quantitative for the most part in the case of partial contact, which is obtained at large sliding velocities $v$ and/or low reduced pressures $p^*$.
The locations of the maxima, $v_\textrm{max}$, are well reflected in the theory, at least on a logarithmic scale. 
Differences between computed and predicted $v_\textrm{max}$ typically approach a factor of two at small reduced pressures. 
The largest discrepancies between theory and simulation occur when the pressure is small, the sliding velocity large, and inertial effects as strong as in regular GFMD.
In that case, theory underestimates the maximum friction coefficient by no more than 50\%.
In contrast, theory overestimates the maximum friction coefficient for SLS and MW-GFMD dynamics.

We continue our comparison between GFMD simulations and Persson's theory in Fig.~\ref{fig:interactionModels} by analyzing how different interactions models affect the dependence of the friction coefficient $\mu \equiv F/L$ on sliding velocity $v$.
In addition to the default interfacial stiffness, $\kappa_\textrm{i}^\textrm{D}$, softer and stiffer overlap penalties are considered.
The theory reflects quite accurately how dissipation increases with increasing interfacial stiffness. 
It also reveals quite clearly that $v_\textrm{max}$ is insensitive to the precise value of $\kappa_\textrm{i}$.
As before, agreement between theory and simulation is better at large than at small normal stress. 

\begin{figure}
    \centering
    \includegraphics[width=0.475\textwidth]{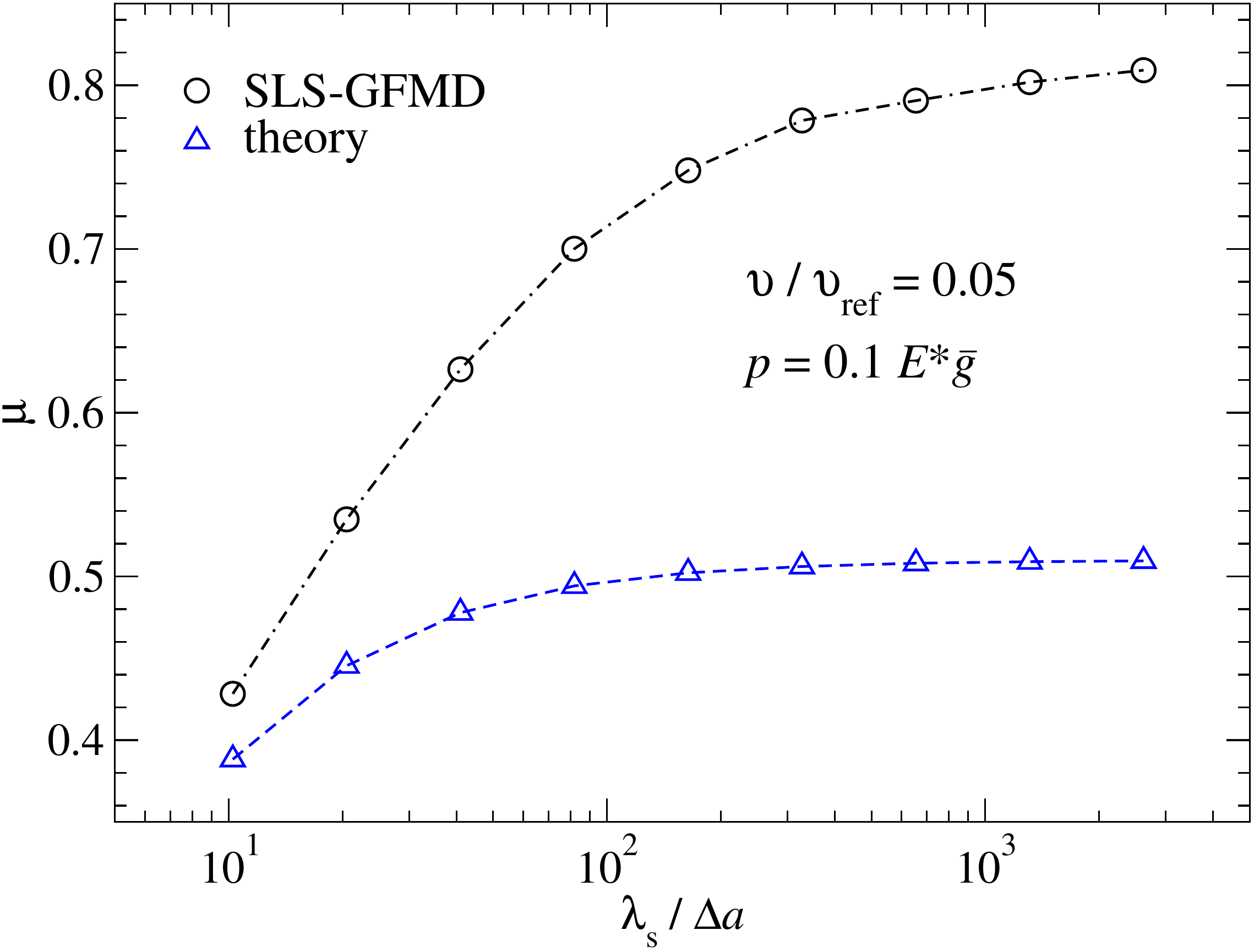}
    \caption{\label{fig:convergence}
     Dependence of the friction coefficient $\mu$ on the discretization $\lambda_\textrm{s}/\Delta a$ at fixed velocity $v/v_{\rm ref} = 0.05$ and fixed reduced pressure of $p^* = 0.1$ for SLS-GFMD (black circles) and Persson's theory (blue triangles).
     The geometry of the rough substrate corresponds to $H=0.8$, $\lambda_{\rm s}/L = 0.04$ and $\lambda_{\rm r}/L = 0.4$.
     Substrate-slider interactions are characterized by $\kappa_\textrm{i} = 0.2\,q_\textrm{max}\,E_2$.
    }
\end{figure}

In our interaction model between slider and elastomer, we consider the overlap penalty to be proportional to the $q_{\rm max}$, which means  the smaller the linear bin size $\Delta a$,
the higher the interfacial stiffness $\kappa_{\rm i}$.
An interesting question to ask is how do theory and simulations converge to the continuum or hard-wall limit? 
To find an answer to this question, we set the sliding velocity and the reduced pressure to constant values, $v/v_{\rm ref} = 0.05$ and $p^* = 0.1$, respectively, and increase the number of grid points in the system, so that the ratio of $\lambda_\textrm{s}$ and the bin size decreases. 
To quickly approach the continuum limit, we consider one-dimensional interfaces, in which case $C(q) \propto q^{-1-2H}$. 
Fig.~\ref{fig:convergence} reveals that the theory deviates more strongly from the simulation results as the continuum limit is reached. 
We rationalize this observation as follows:
The theory does not assume displacement modes with wave vectors $q > q_\textrm{s}$ so that displacement modes $\tilde{u}(q>q_\textrm{s})$ cannot dissipate energy.
However, they do in ``reality'', unless they do not exist due to the absence of sub-atomic atoms. 
Since self-affine roughness is frequently observed down to the smallest scales, we would argue that studying the approach to the continuum limit is somewhat of a predominantly mathematical exercise.
Yet, what can be learned from it is that the coupling of static or time-dependent $\tilde{h}(q)$ modes to modes associated with larger wave vectors does contribute to the overall dissipation and that the coupling increases with the stiffness of the interactions between rigid, rough slider and elastomer. 

For our final comparisons between Persson's theory and GFMD simulations, we varied the surface spectra.
One time, we only changed the way how $C(q))$ crosses over from the self-affine branch to the small-wave-number domain, which were, cut-off as well as abrupt and smooth roll-off.
Effects on the friction are marginal at small velocities. 
However, at high velocities we observe some differences between the three choices.
This is because dissipation at large velocities is related to longer wavelength undulations, which is where the three approaches differ. 
The more dramatic change of surface spectrum was the substitution of the Hurst exponent from $H = 0.8$ to $H = 0.3$. 
Relative effects are again reproduced quite closely. 
Relative errors in the friction force for $H = 0.3$ are slightly larger at small velocities than for $H = 0.8$, however, the overall trends are matched again quite accurately. 

\begin{figure*}
    \centering
    \includegraphics[width=0.32\textwidth]{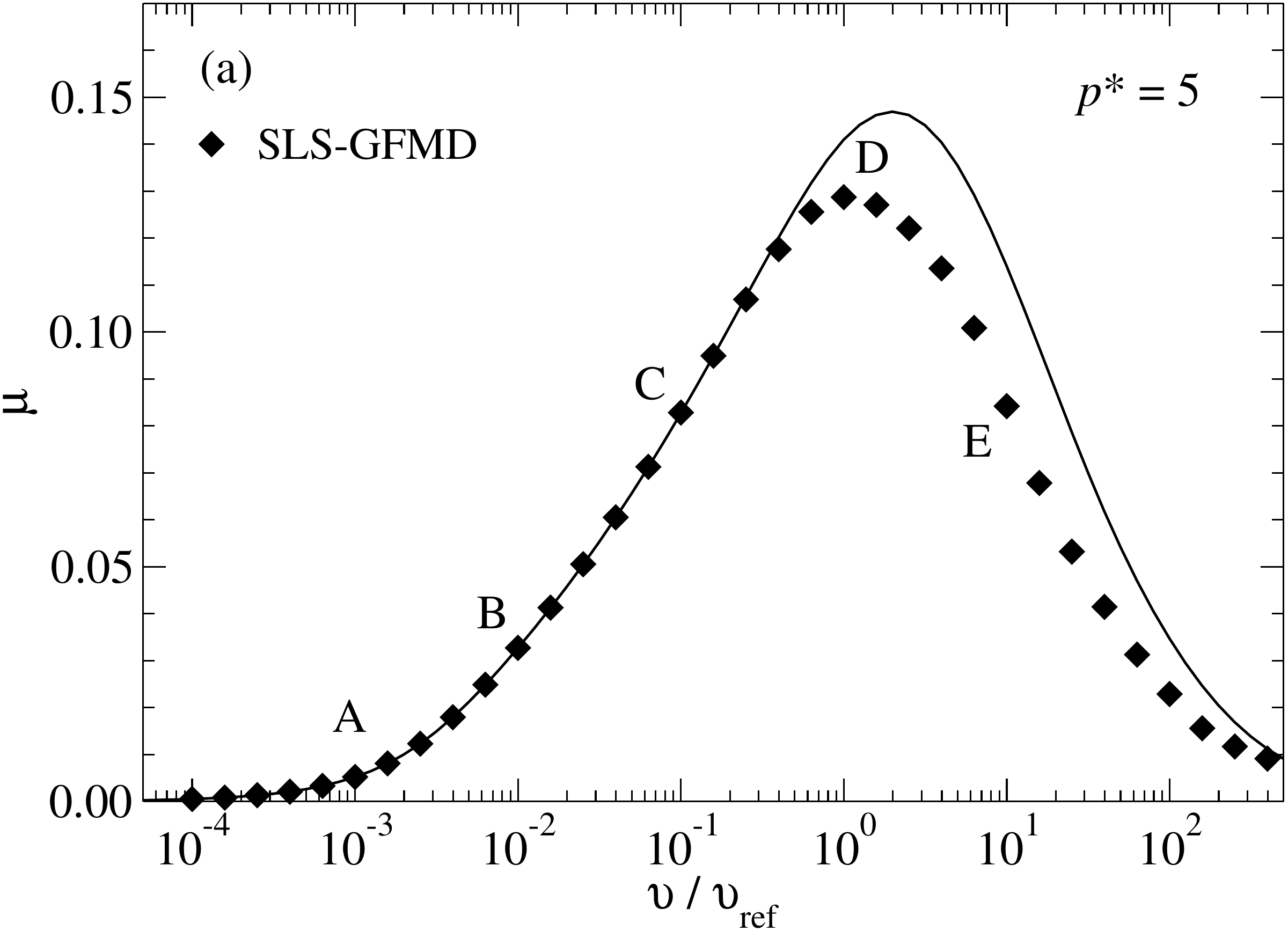}
    \includegraphics[width=0.32\textwidth]{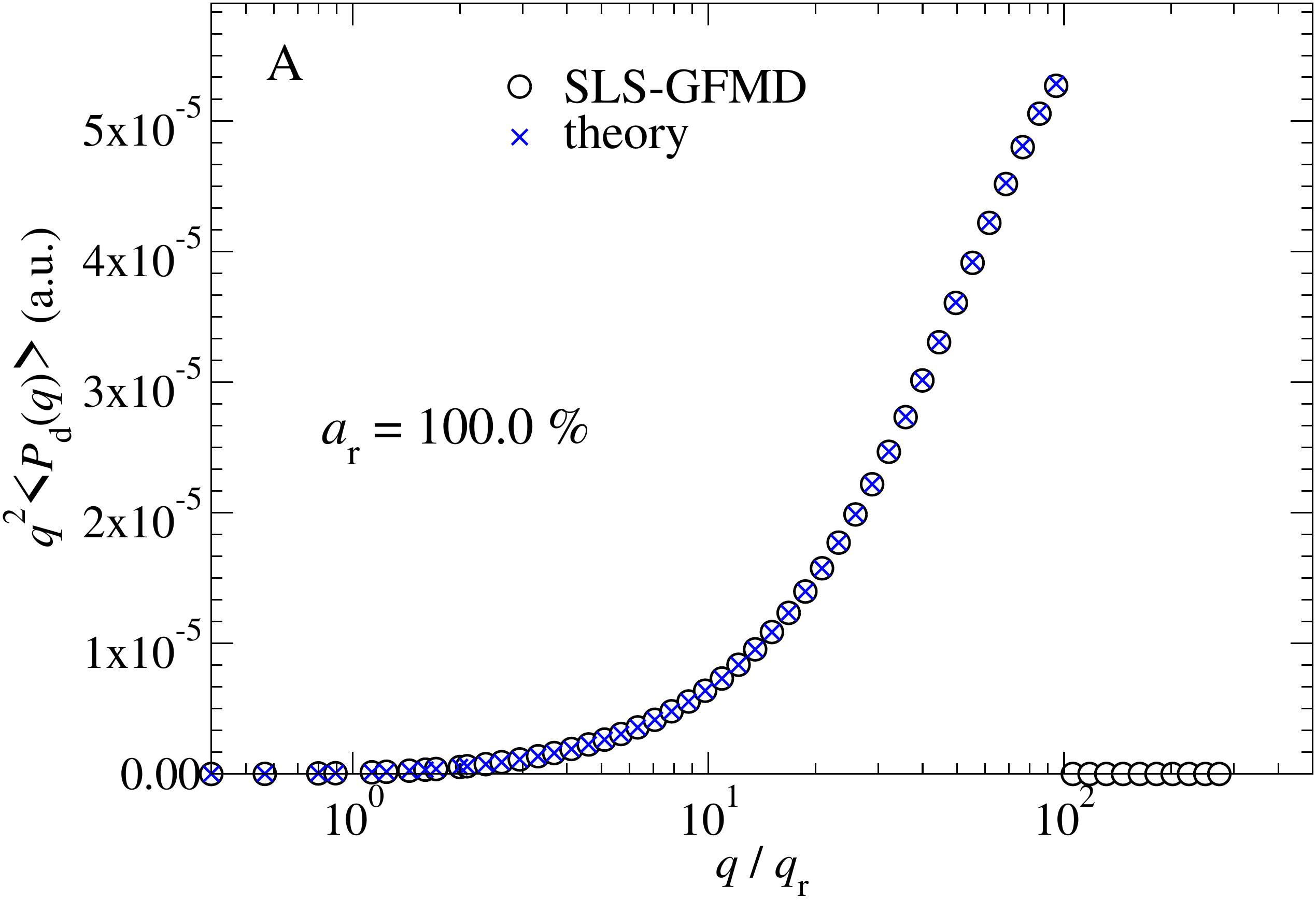}
    \includegraphics[width=0.32\textwidth]{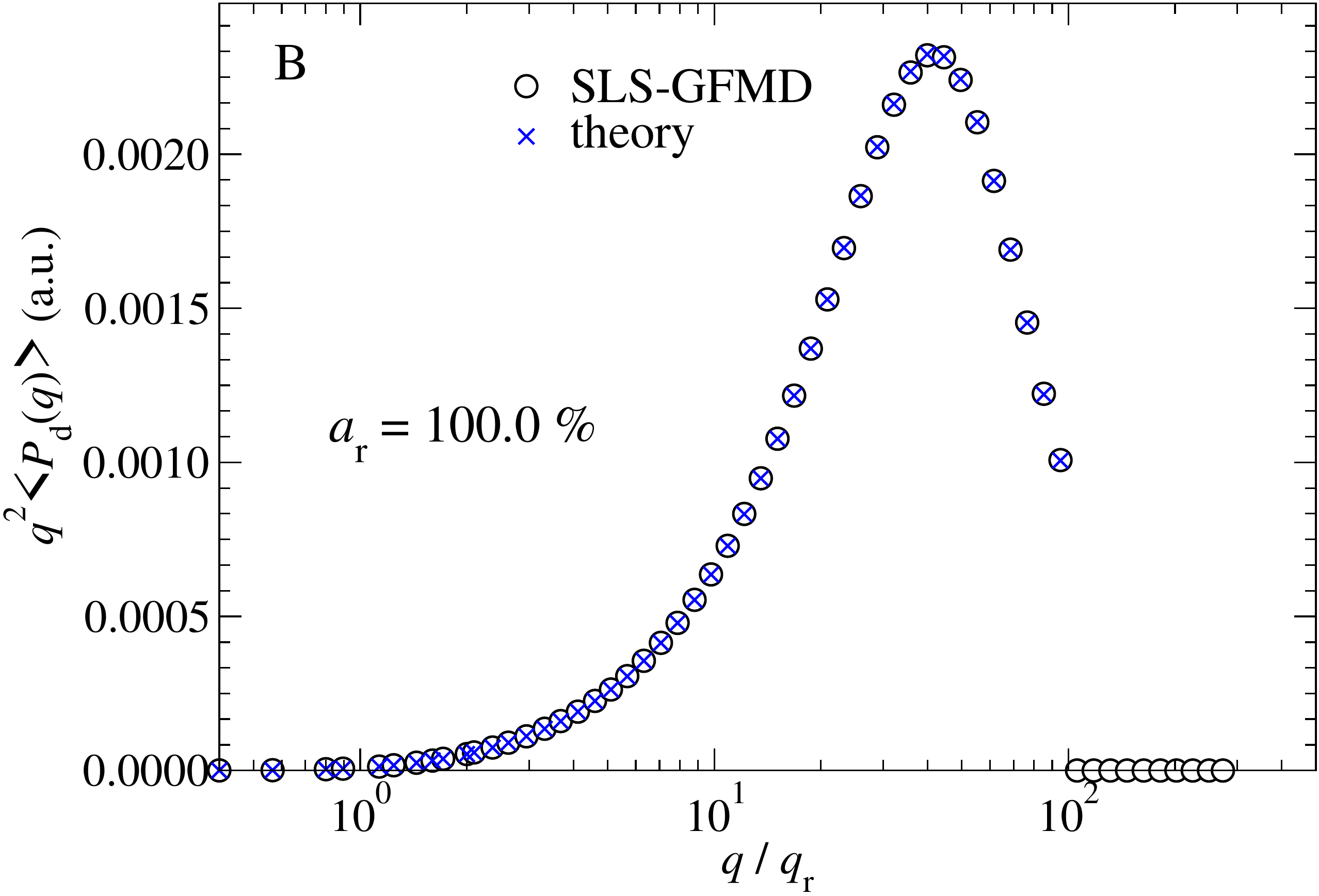}
    \includegraphics[width=0.32\textwidth]{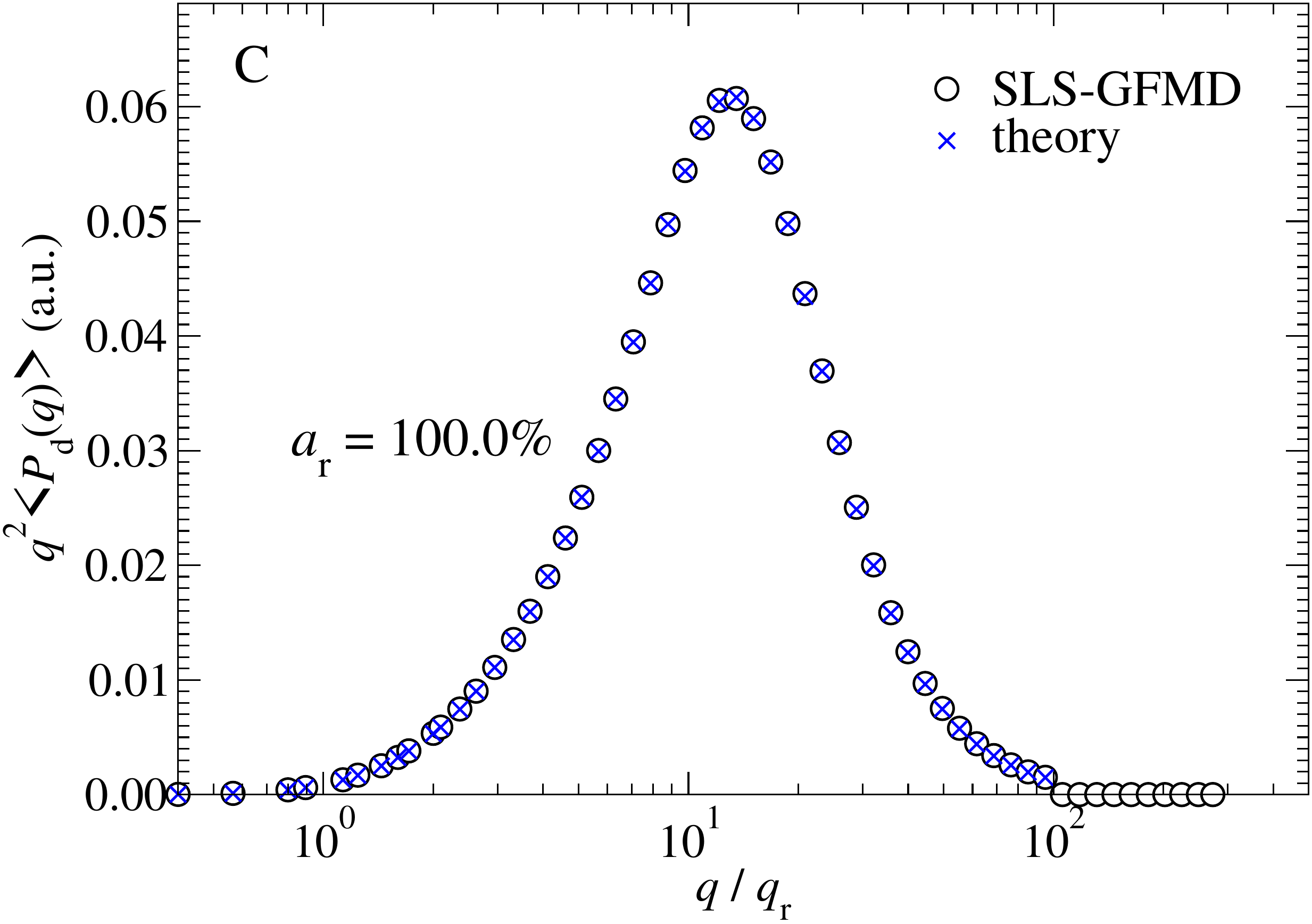}
    \includegraphics[width=0.32\textwidth]{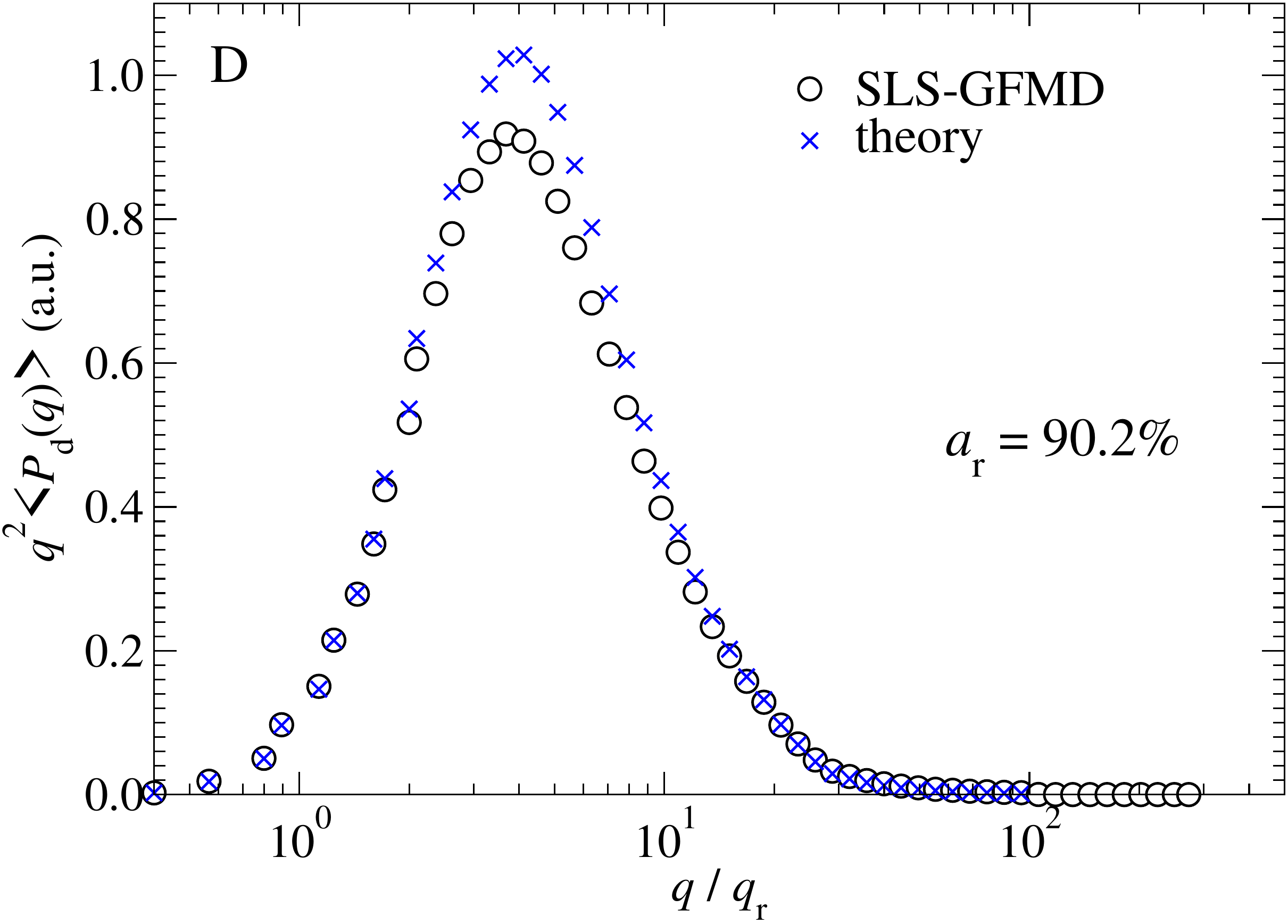}
    \includegraphics[width=0.32\textwidth]{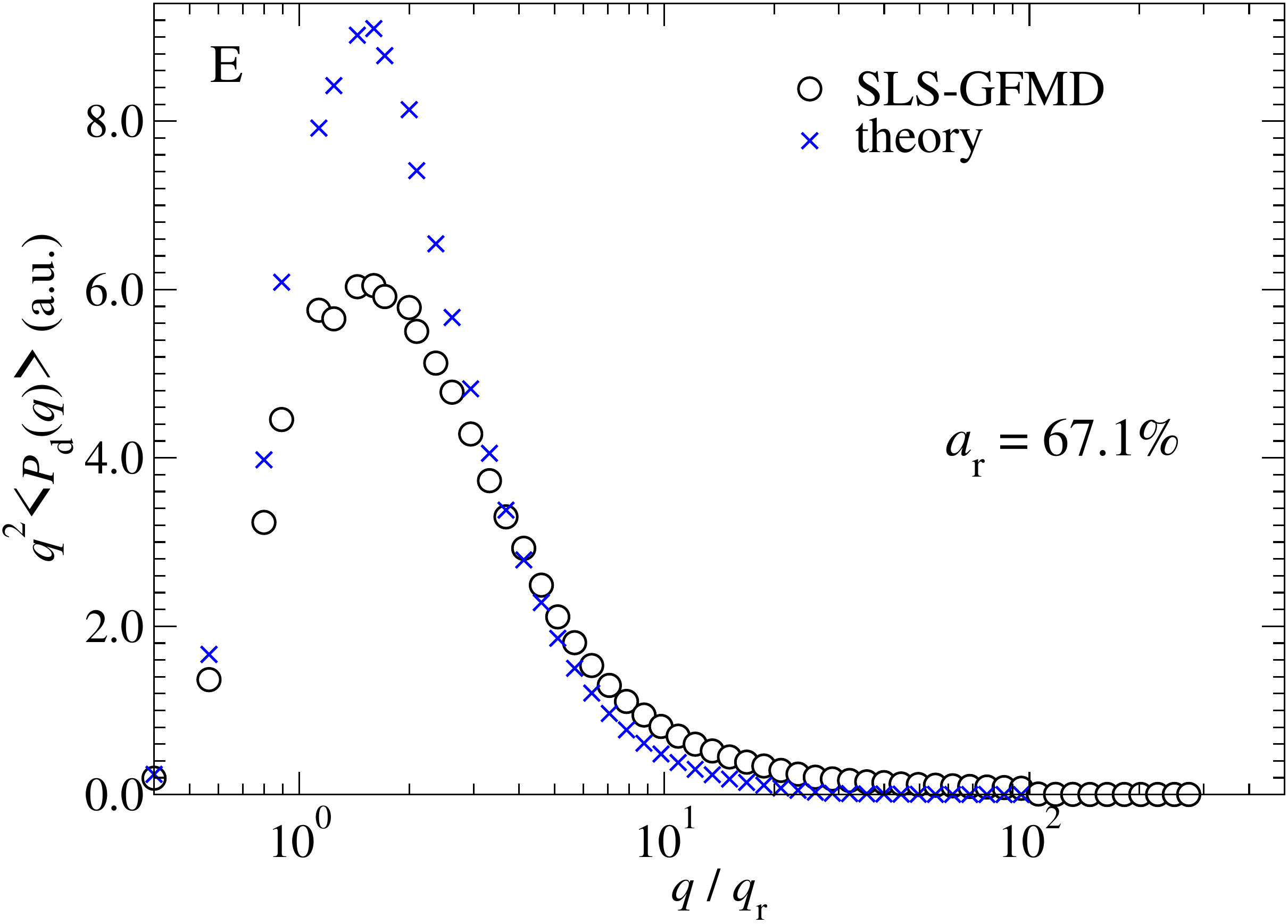}
    \caption{(a) Friction coefficient $\mu$ as a function of velocity $v$ at a pressure of $p^* = 5$ for SLS-GFMD (black diamonds) and theory (full line), and the expectation value of the dissipated power $P_{\rm{d}}(q, v)$ as a function of $q$ for several fixed values of velocities: (A) $v/v_{\rm{ref}}=10^{-3}$, (B) $v/v_{\rm{ref}}=10^{-2}$, (C) $v/v_{\rm{ref}}=10^{-1}$, (D) $v/v_{\rm{ref}}=10^{0}$, and (E) $v/v_{\rm{ref}}=10^{1}$. SLS-GFMD results are drawn as black circles, theoretical predictions - as blue crosses. The parameters of the rigid indenter are set to default options.}
    \label{fig:dissQhighP}
\end{figure*}

To better rationalize the discrepancies between theory and simulations, we conducted an additional analysis, in which we resolved the dissipated power as a function of wave number $q$ and velocity for the default set-up. 
This was done by computing the function $<P_{\rm{d}}(q,v)>$, which represents the expectation value of individual summands on the r.h.s. of Eq.~\eqref{eq:disspatedPowerTimeDomainB}. 
In the limit of  high pressures and low velocities, the theory is exact, as is clearly borne out in Fig.~\ref{fig:dissQhighP}(A-C).
This is because full contact is reached and the theory assumes the full-contact solution as input.
However, when contact starts being partial, discrepancies appear. 
Yet, at high pressure, the overall shape of the curves including the location of maxima are well matched, as can be seen in Fig.~\ref{fig:dissQhighP}(D-E).

\begin{figure*}
    \centering
    \includegraphics[width=0.32\textwidth]{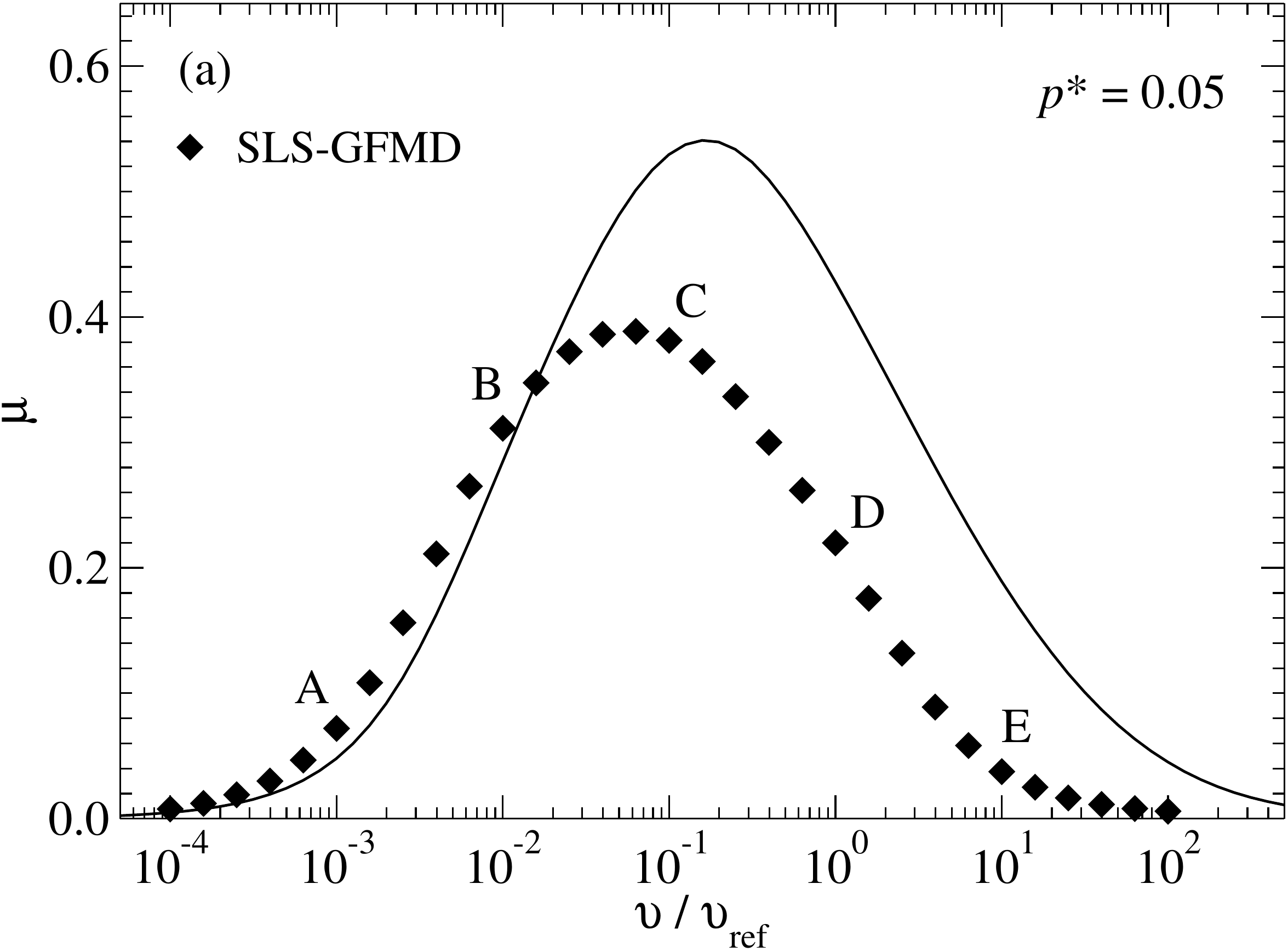}
    \includegraphics[width=0.32\textwidth]{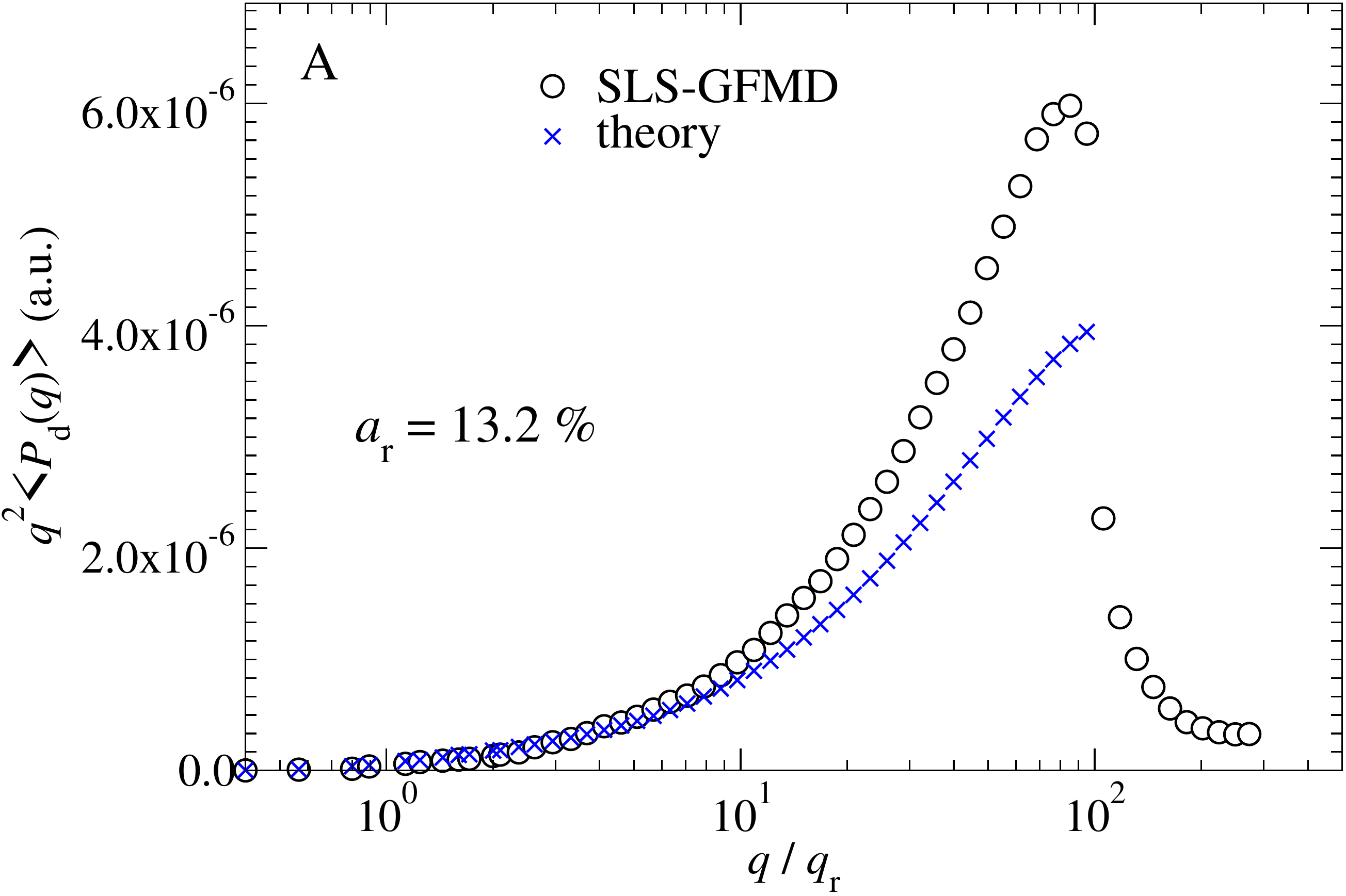}
    \includegraphics[width=0.32\textwidth]{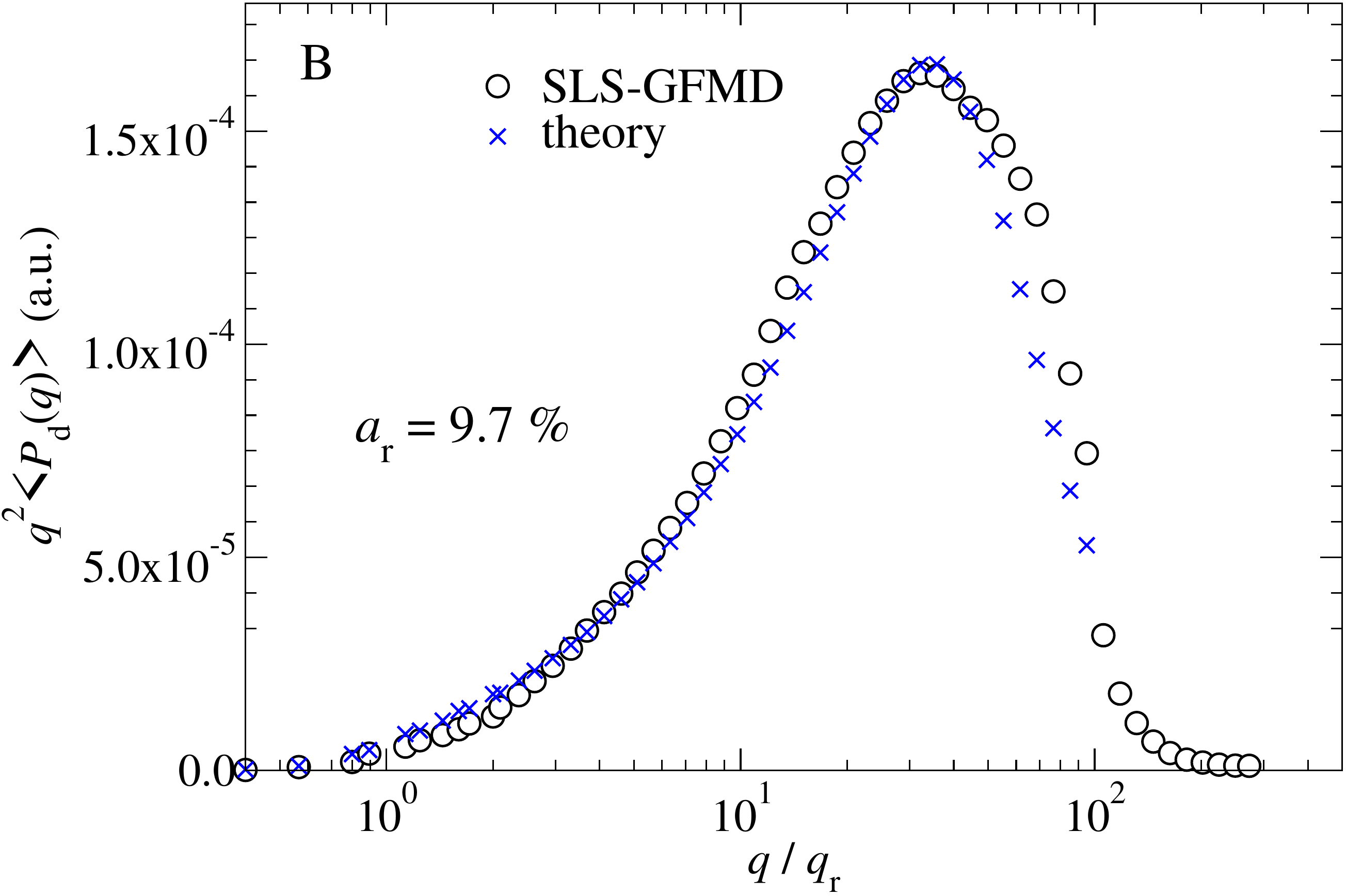}
    \includegraphics[width=0.32\textwidth]{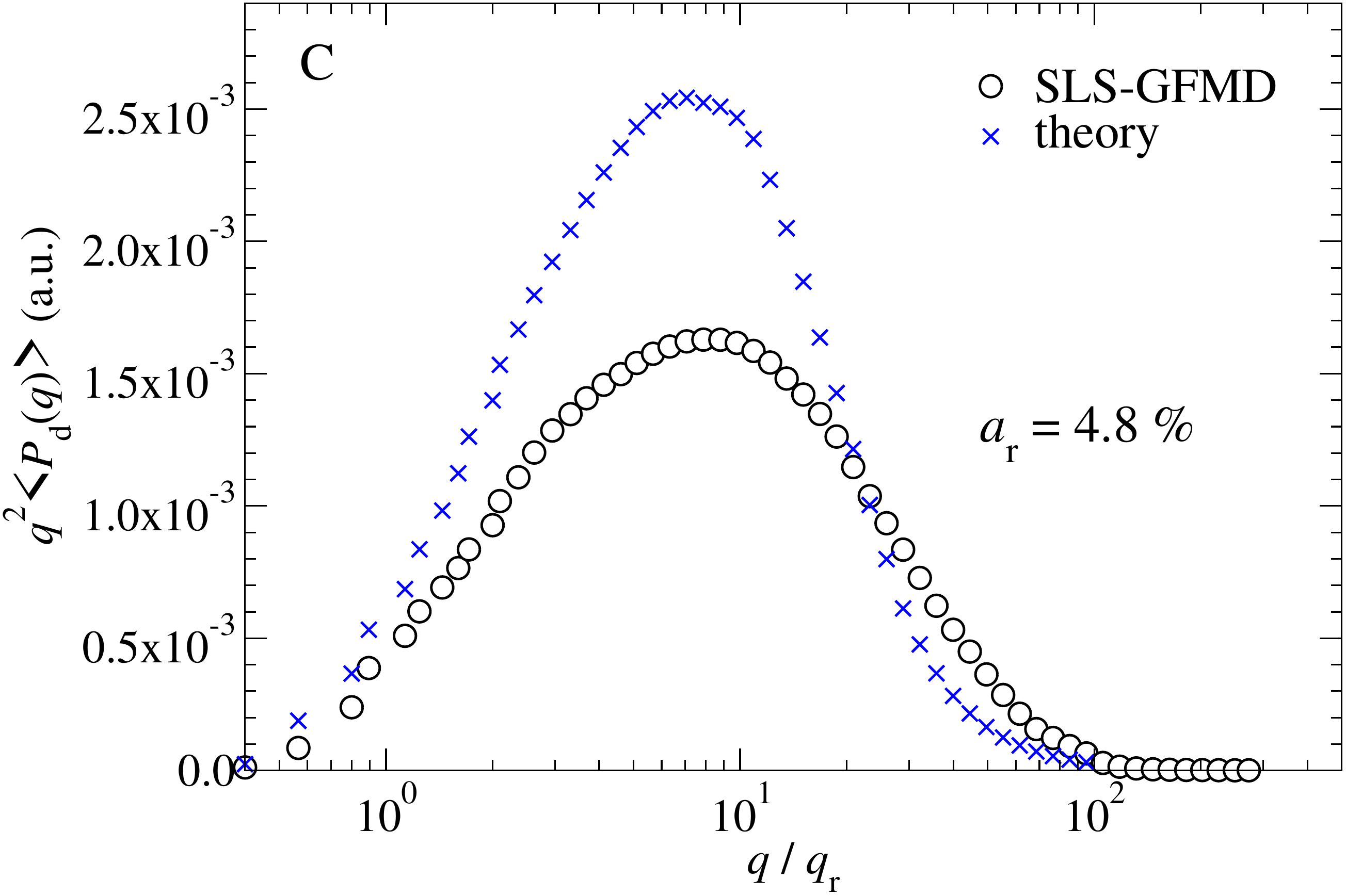}
    \includegraphics[width=0.32\textwidth]{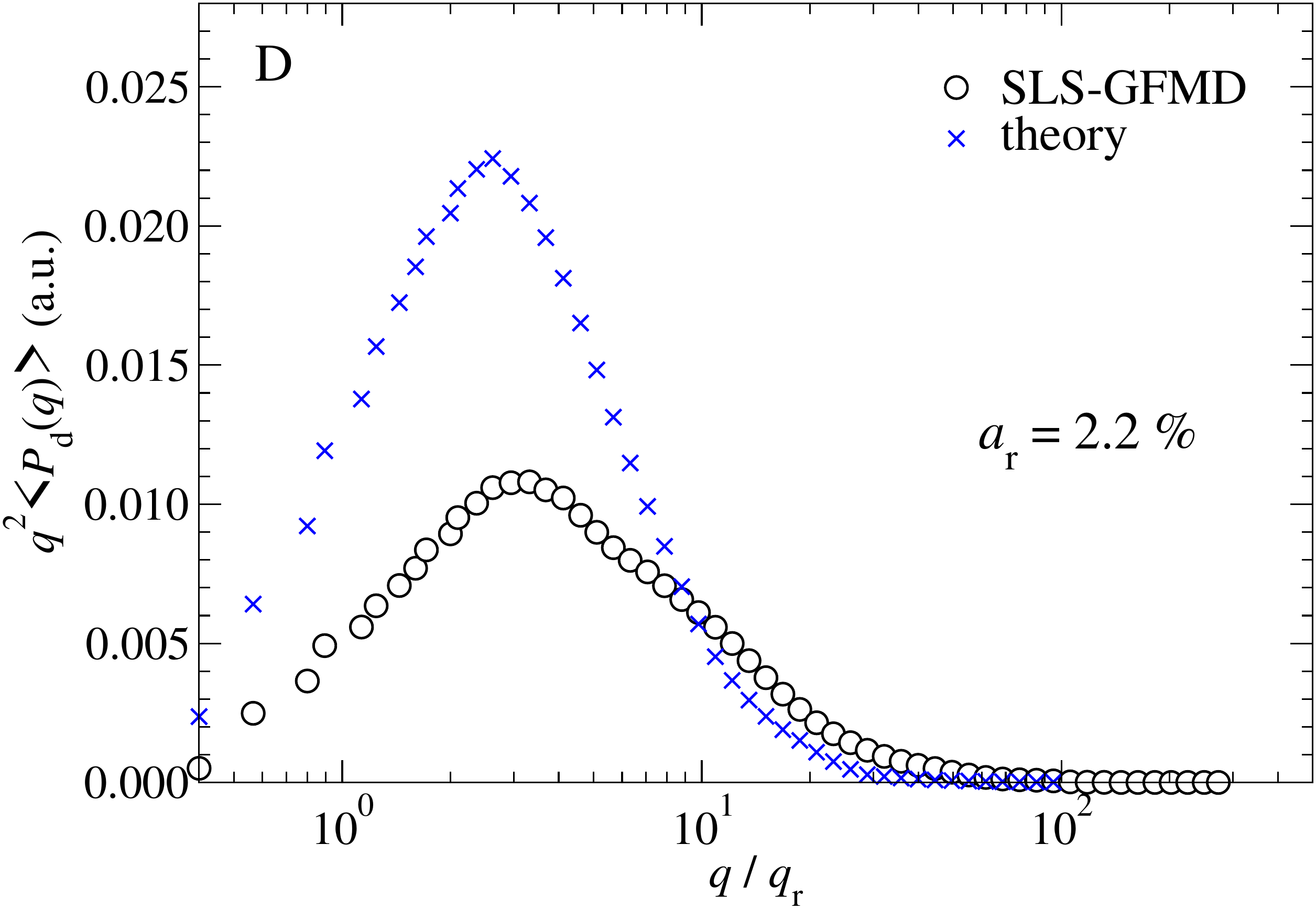}
    \includegraphics[width=0.32\textwidth]{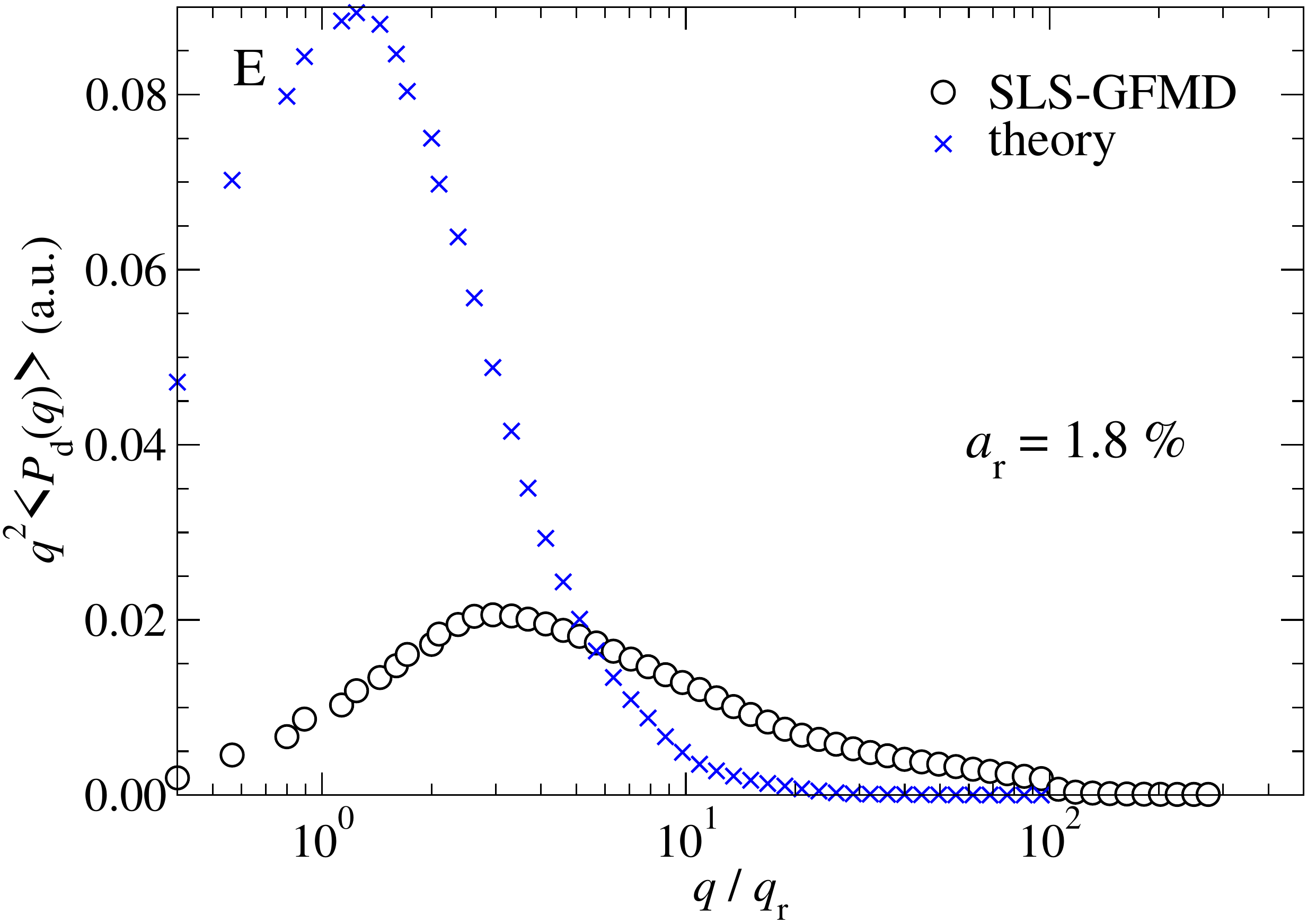}
    \caption{Same as Fig.~\ref{fig:dissQhighP} but for lower pressure of $p^* = 0.05$.}
    \label{fig:dissQlowP}
\end{figure*}

For low pressures, e.g., one leading to approximately 10~\% contact area at $v = 0$, the theory underestimates the friction coefficient at low speeds, as was revealed in Figs.~\ref{fig:interactionModels}--\ref{fig:substrateModels}.
The discrepancy originates partly from the deformations that occur at wavelengths smaller than $\lambda_\textrm{s}$, which are neglected in the theory.
The contribution of these modes are revealed most clearly by the tails of the simulation data for $q/q_\textrm{r}>100$ in Fig.~\ref{fig:dissQlowP} for state points A and B. 
Similarly, the theory ignores that an undulation at wave vector $q$ can excite, for example, a time-dependent and thus dissipative undulation with wave vector $2q$, even if $2q$ is less than $2q_\textrm{s}$.

Significant discrepancies occur at large velocities and small pressure.
In that limit, the shape of the theoretical and simulated $P_{\rm{d}}(q,v)$ differ and moreoever, height and location of the maxima are substantially shifted with respect to each other. %
However, except for large $v$ and small $p$, the $q$-dependence of the dissipated power is predicted quite well by the theory in steady-state sliding.

\section{Conclusions}
\label{sec:conclusions}

In this work, we used GFMD to study various contact models with the goal to ascertain the validity of Persson's contact mechanics theory for \emph{sliding} interfaces between an elastomer and a rigid, randomly rough counterface.
The theory reproduces at least semi-quantitatively how friction increases with sliding velocity $v$ up to the point of maximum friction.
At larger sliding velocities, i.e., once sliding is so fast that it reduces noticeably the true contact area, agreement between theory and simulation remains qualitative. 
However, relative errors in the predicted and the computed friction force can be significant at large $v$.
This weakness of the theory is observed for all studied dynamical systems but is most pronounced when inertial effects become important. 
One reason for the discrepancy between theory and simulation is that relaxation and thus dissipation occurs more and more outside of the contact close to its trailing edge as velocity increases, while the friction force acts increasingly at the leading edge of contact, similar to what has been observed in the atomistic simulation of surfactant molecules~\cite{Gao2021L}.
However, the theory assumes dissipation to occur symmetrically in the contact, i.e., as much at the leading as at the trailing edge, as is, in fact, the case at small sliding velocities.
It does not include the sliding-induced asymmetry in displacement and stresses.
Unfortunately, it is not clear to us yet how to encode this insight into Persson's description of sliding contacts.
However, it may not be particularly important to do so, because absolute corrections are small. 

In conclusion, Persson's approach to time-dependent, self-affine contacts reproduces the observed effects on the friction coefficient at least semi-quantitatively, except for highly inertial systems at small pressure and medium to large sliding velocities. 
Successful predictions include the way how maximum friction coefficient $\mu_\textrm{max}$ and their location $v_\textrm{max}$, change with the parameters defining the model.
In general, the absolute values for $\mu_\textrm{max}$ have errors of less than 50\%, while $v_\textrm{max}$ tends to be underestimated by a factor of two at small reduced pressures.
Given the simplicity of the theory, this level of agreement can only be deemed remarkable.

\noindent
\textbf{Acknowledgements.} Both authors acknowledge helpful discussions with Bo Persson and Michele Scaraggi. This work was supported by the DFG through grant MU 1694/5-2. MHM is grateful for support through an INM Fellowship.  \vspace*{1mm} \\
\textbf{Declaration of Interests.} The authors declare that they have no known competing financial interests or personal relationships that could have appeared to influence the work reported in this paper.

\bibliographystyle{ieeetr}
\bibliography{main}
\end{document}